\documentclass[a4paper,twoside]{cpc-hepnp}
\usepackage{multicol}
\usepackage{lastpage}
\usepackage{booktabs}
\usepackage{ragged2e}

\usepackage{graphicx}
\usepackage{bm}
\usepackage[tbtags]{amsmath}
\usepackage[utf8]{inputenc}
\usepackage{amsfonts}
\usepackage{amssymb}
\usepackage{multirow}
 \usepackage[colorlinks,linktocpage,linkcolor=cyan,citecolor=cyan]{hyperref}
\usepackage{adjustbox}
\usepackage{array}
\usepackage[capitalise]{cleveref}
\usepackage{acro}
\usepackage[utf8]{inputenc}
\usepackage{amssymb}
\usepackage{subfigure}

\usepackage{tikz}
\usepackage{pgfplots}
\pgfplotsset{compat=newest,every axis plot/.append style={line width=1pt}}

\crefname{figure}{Fig.}{Figs.}
\Crefname{figure}{Fig.}{Figs.}
\def\({\left(}
\def\){\right)}
\def\[{\left[}
\def\]{\right]}

\newcommand{\be}{{\begin{eqnarray}}}
\newcommand{\ee}{{\end{eqnarray}}}

\newcommand{\Beq}{\begin{align}}
\newcommand{\Eeq}{\end{align}}

\DeclareAcronym{SW}{
	short = SW ,
	long = Sachs-Wolfe ,
	short-plural =  ,
}
\DeclareAcronym{BH}{
	short = BH ,
	long = black hole ,
	short-plural = s ,
}
\DeclareAcronym{SNR}{
	short = SNR ,
	long = signal-to-noise ratio ,
	short-plural = s ,
}
\DeclareAcronym{IMRPPv2}{
	short = ,
	long = {\normalsize IMRP}{\footnotesize HENOM}{\normalsize P}v2 ,
	short-plural = ,
}
\DeclareAcronym{SFR}{
	short = SFR ,
	long = star formation rate ,
	short-plural =  ,
}
\DeclareAcronym{IMR}{
	short = IMR ,
	long = inspiral-merger-ringdown ,
	short-plural =  ,
}
\DeclareAcronym{ABH}{
	short = ABH ,
	long  = astrophysical black hole,
	short-plural = s ,
}
\DeclareAcronym{GW}{
	short = GW ,
	long = gravitational wave ,
	short-plural = s ,
}
\DeclareAcronym{SIGW}{
	short = SIGW ,
	long = scalar-induced gravitational wave ,
	short-plural = s ,
}
\DeclareAcronym{GWB}{
	short = GWB ,
	long = gravitational-wave background ,
	short-plural = s ,
}
\DeclareAcronym{CBC}{
	short = CBC ,
	long = compact binary coalescence ,
	short-plural = s ,
}
\DeclareAcronym{BBH}{
	short = BBH ,
	long = binary black hole ,
	short-plural = s ,
}
\DeclareAcronym{PBH}{
	short = PBH ,
	long = primordial black hole ,
	short-plural = s ,
}
\DeclareAcronym{LIGO}{
	short =LIGO ,
	long = Laser Interferometer Gravitational-Wave Observatory ,
	short-plural = ,
}
\DeclareAcronym{LVK}{
	short = LVK ,
	long = {Advanced LIGO, Virgo and KAGRA} ,
	short-plural = ,
}
\DeclareAcronym{ET}{
	short = ET ,
	long  = Einstein Telescope,
	short-plural =  ,
}
\DeclareAcronym{CE}{
	short = CE ,
	long  = Cosmic Explorer,
	short-plural =  ,
}
\DeclareAcronym{LISA}{
	short = LISA ,
	long  = Laser Interferometer Space Antenna,
	short-plural =  ,
}
\DeclareAcronym{BBO}{
	short = BBO ,
	long  = big bang observer,
	short-plural =  ,
}
\DeclareAcronym{DECIGO}{
	short = DECIGO ,
	long  = Deci-hertz Interferometer Gravitational wave Observatory,
	short-plural =  ,
}
\DeclareAcronym{SKA}{
	short = SKA ,
	long  = Square Kilometre Array,
	short-plural =  ,
}
\DeclareAcronym{PTA}{
	short = PTA ,
	long = pulsar timing array ,
	short-plural = s ,
}
\DeclareAcronym{FRW}{
	short = FRW ,
	long = Friedmann-Robertson-Walker ,
	short-plural =  ,
}
\DeclareAcronym{CMB}{
	short = CMB ,
	long = cosmic microwave background ,
	short-plural =  ,
}
\DeclareAcronym{RD}{
	short = RD,
	long  = radiation-dominated ,
	short-plural =  ,
}
\DeclareAcronym{MD}{
	short = MD,
	long  = matter-dominated ,
	short-plural =  ,
}
\DeclareAcronym{HD}{
	short = HD,
	long  = Hellings-Downs ,
	short-plural =  ,
}
\DeclareAcronym{SMBH}{
	short = SMBH ,
	long  = supper-massive black hole ,
	short-plural = s ,
}
\DeclareAcronym{SGWB}{
	short = SGWB ,
	long  = stochastic gravitational-wave background ,
	short-plural = s ,
}
\DeclareAcronym{NG}{
	short = NANOGrav ,
	long  = North American Nanohertz Observatory for Gravitational Waves ,
	short-plural =  ,
}
\DeclareAcronym{PSD}{
	short = PSD ,
	long  = power spectral density ,
	short-plural = s ,
}
\DeclareAcronym{PDF}{
	short = PDF ,
	long  = probability distribution function ,
	short-plural = s ,
}
\DeclareAcronym{BBN}{
	short = BBN ,
	long  = big-bang nucleosynthesis ,
	short-plural =  ,
}
\DeclareAcronym{EoS}{
	short = EoS ,
	long  = equation of state ,
	short-plural =  ,
}

\begin{document}

\fancyhead[c]{\small Chinese Physics C~~~Vol. 37, No. 1 (2013) 010201}
\fancyfoot[C]{\small010201-\thepage}
\footnotetext[0]{Received 14 March 2009}

\title{
Unraveling the early universe's equation of state and primordial black hole production with PTA, BBN, and CMB observations }

\author{
    Qing-Hua Zhu$^1$
    \quad Zhi-Chao Zhao$^2$
     \quad Wang Sai$^3$ \email{wangsai@ihep.ac.cn (Corresponding author)}
      \quad Xin Zhang$^{4,5,6}$ \email{zhangxin@mail.neu.edu.cn (Corresponding author)}
}
\maketitle
\address{
$^1$ Department of Physics and Chongqing Key Laboratory for Strongly Coupled Physics, Chongqing University, Chongqing 401331, People's Republic of China\\
$^2$ Department of Applied Physics, College of Science, China Agricultural University,	Qinghua East Road,  Beijing 100083, People's Republic of China\\
$^3$ Theoretical Physics Division, Institute of High Energy Physics, Chinese Academy of Sciences,  Beijing 100049, People's Republic of China\\
$^4$ Key Laboratory of Cosmology and Astrophysics (Liaoning) \& College of Sciences, Northeastern University,  Shenyang 110819, People's Republic of China\\
$^5$ National Frontiers Science Center for Industrial Intelligence and Systems Optimization, Northeastern University,  Shenyang 110819, People's Republic of China\\
$^6$ Key Laboratory of Data Analytics and Optimization for Smart Industry (Ministry of Education), Northeastern University,  Shenyang 110819, People's Republic of China\\
}
	
	

\begin{abstract}
Pulsar timing array (PTA) data releases showed strong evidence for a stochastic gravitational-wave background in the nanohertz band. When the signal is interpreted by a scenario of scalar-induced gravitational waves (SIGWs), we encounter overproduction of primordial black holes (PBHs). We wonder if varying the equation of state (EoS) of the early Universe can resolve this issue and thereby lead to a consistent interpretation of the PTA data. Analyzing a data combination of PTA, big-bang nucleosynthesis, and cosmic microwave background, we find that an epoch with EoS $w\sim\mathcal{O}(10^{-2})$ between the end of inflation and the onset of radiation domination can significantly suppress the production of PBHs, leading to alleviation of the PBH-overproduction issue. With the inferred interval $w=0.44_{-0.40}^{+0.52}$ at 95\% confidence level, our scenario can interpret the PTA data just as well as the conventional scenario of SIGWs produced during the radiation domination. 
\end{abstract}


\footnotetext[0]{\hspace*{-3mm}\raisebox{0.3ex}{$\scriptstyle\copyright$}2013
	Chinese Physical Society and the Institute of High Energy Physics
	of the Chinese Academy of Sciences and the Institute
	of Modern Physics of the Chinese Academy of Sciences and IOP Publishing Ltd}%

\begin{multicols}{2}


\section{Introduction}

Recently, multiple data releases of \acp{PTA} showed significant evidence for a \acl{SGWB} in the nanohertz band \cite{Xu:2023wog,EPTA:2023fyk,NANOGrav:2023gor,Reardon:2023gzh}. 
The signal has been interpreted to arise from \acp{SIGW} by the authors of Refs.~\cite{Franciolini:2023pbf,Inomata:2023zup,Cai:2023dls,Wang:2023ost,Liu:2023ymk,Abe:2023yrw,Ebadi:2023xhq,Figueroa:2023zhu,Yi:2023mbm,Madge:2023dxc,Firouzjahi:2023lzg,Wang:2023sij,You:2023rmn,Ye:2023xyr,HosseiniMansoori:2023mqh,Balaji:2023ehk,Jin:2023wri,Das:2023nmm}. 
Though this scenario fits the \ac{PTA} data better than other scenarios, it results in an issue of overproduction of \acp{PBH} \cite{NANOGrav:2023hvm,EPTA:2023xxk}. 
In these works, the authors have assumed that \acp{SIGW} were produced during an epoch of radiation domination with the \ac{EoS} $w=1/3$. 
However, the \ac{EoS} of the early Universe might deviate from the radiation domination \cite{Drees:2015exa,Saikawa:2018rcs,Carr:2019kxo,Dalianis:2020gup,Domenech:2020ssp,Lozanov:2022yoy,Bhaumik:2020dor,Haque:2021dha,Papanikolaou:2020qtd,Huang:2023chx,Johnson:2008se,Turner:1983he,Poulin:2018dzj,Vikman:2004dc,Domenech:2021wkk}. 
We wonder if the overproduction issue can be resolved via considering an arbitrary \ac{EoS} of the early Universe during an epoch between the end of inflation and the onset of radiation domination.

It is well known that \acp{SIGW} could be nonlinearly produced by linear cosmological curvature perturbations, when the latter reentered the Hubble horizon after the end of inflation \cite{Ananda:2006af,Baumann:2007zm,Mollerach:2003nq,Assadullahi:2009jc,Espinosa:2018eve,Kohri:2018awv}. 
The energy-density spectrum of \acp{SIGW} would be significantly affected by the \ac{EoS} of the Universe at the production time of \acp{SIGW} \cite{Kohri:2018awv,Inomata:2020tkl,Inomata:2019ivs,Hajkarim:2019nbx,Domenech:2019quo,Domenech:2020kqm,Domenech:2020ers,Domenech:2021wkk,Domenech:2021ztg,Dalianis:2020gup}. 
The corresponding imprints may be left on the \ac{PTA} signal, since the Universe is transparent to the gravitational waves \cite{Flauger:2019cam}. 
Therefore, we anticipate that \acp{SIGW} is an effective probe of the early-Universe physics, particularly for the \ac{EoS} parameter.

It is also known that \acp{PBH} were produced due to gravitational collapse of these perturbations if the latter excess a critical overdensity when they reentered into the Hubble horizon \cite{Carr:1974nx}. 
The mass function of \acp{PBH}, and thereby the abundance, would also be significantly impacted by the \ac{EoS} of the Universe at the formation time of \acp{PBH} \cite{Domenech:2020ers,Harada:2016mhb,Nakama:2020kdc,Inomata:2020lmk}.
Therefore, the issue of \ac{PBH} overproduction might be resolved when an appropriate \ac{EoS} of the early Universe is taken into account.

In this work, we study a scenario of \acp{SIGW} produced during the epoch with an arbitrary \ac{EoS} between the end of inflation and the onset of radiation domination. 
Differing from studies assuming that all of \acp{SIGW} were produced during the $w$-domination before the onset of radiation domination \cite{Domenech:2019quo,Domenech:2021ztg}, we consider that \acp{SIGW} were produced during both epochs. 
Our scenario is inspired by an enhancement mechanism for \acp{SIGW} caused by a sudden transition from the early-matter domination to the radiation domination \cite{Inomata:2020tkl,Inomata:2019ivs}. 
We expect that a similar enhancement mechanism can also exist in our scenario. 
By fitting the \ac{NG} 15-year data release \cite{NANOGrav:2023gor} to our scenario, we study whether varying \ac{EoS} of the early Universe can improve the ability of \acp{SIGW} to interpret the \ac{PTA} signal. 
Meanwhile, we propose that the enhancement of \ac{SIGW} spectrum can be significantly constrained by further taking into account the \ac{BBN} \cite{Cooke:2013cba} and \ac{CMB} \cite{Clarke:2020bil} data. 
It is worth noting that our results would be different from those of Refs.~\cite{NANOGrav:2023hvm,EPTA:2023xxk}, where the authors considered \acp{SIGW} produced during the radiation domination only and analyzed the \ac{PTA} data only. 
Moreover, we study the production process of \acp{PBH} during the $w$-dominated epoch. 
We will show that the \ac{PBH} abundance is significantly changed, probably leading to a resolution of the \ac{PBH}-overproduction issue.

The remaining context of this paper is arranged as follows.  
In \cref{sec:theory}, we develop the scenario of \acp{SIGW} by taking into account the arbitrary \ac{EoS} of the early Universe. 
In \cref{sec:method}, we demonstrate the method of data analysis and show the inferred parameter region. 
In \cref{III}, we study the formation of \acp{PBH} in our scenario and the issue of \ac{PBH} overproduction.
In \cref{sec:conclusion}, we show conclusions and discussion.

\section{SIGW energy-density spectrum}\label{sec:theory}

It has been known that a sudden transition from the early-matter domination to the radiation domination can result in a significant enhancement of the energy-density spectrum of \acp{SIGW} \cite{Inomata:2020tkl,Inomata:2019ivs}. 
In this work, we replace the early-matter domination with an epoch dominated by ``matter'' with an arbitrary \acp{EoS} parameter $w \in [0,1]$, following the research method of Refs.~\cite{Hajkarim:2019nbx,Domenech:2019quo, Domenech:2021ztg}. 
However, we still consider the sudden transition, since in general, it is too complicated to study a finite duration of the transition, which is left to future works. 
We will study the imprints of such a scenario on the energy-density spectrum of \acp{SIGW} in the following. 
In fact, there are several mechanisms to generate the sudden transition from the early-matter domination to the radiation domination \cite{Inomata:2020lmk,Domenech:2020ssp,Domenech:2021wkk,Dalianis:2021dbs,Lozanov:2022yoy,White:2021hwi,Kasuya:2022cko,Kawasaki:2023rfx,Flores:2023dgp,Harigaya:2023mhl,Inomata:2019ivs}.

We adopt a sudden transition of the scale factor $a$, i.e., \cite{Inomata:2019ivs} 
\begin{eqnarray}
\frac{a (\eta)}{a_\text{R}}  =  \left\{\begin{array}{ll}
\left( \frac{\eta}{\eta_\text{R}} \right)^{\frac{2}{1 + 3 w}} & \quad\quad  \mathrm{for}~~\eta < \eta_\text{R} \\
\frac{2}{(1+3w)}\left(\frac{\eta}{\eta_\text{R}}-1\right)+1 & \quad\quad  \mathrm{for}~~\eta \geq \eta_\text{R}
\end{array}  \right.  \label{a} ~,
\end{eqnarray}
where $\eta$ is the conformal time. 
We introduce $a_{\text{R}}$ with $\eta_\text{R}$ at the occasion of transition, i.e., 
\begin{eqnarray}
    a_\text{R}&=& a(\eta_{\text{R}})=\left(\frac{1+3w}{2}\right)H_0\eta_\text{R}\sqrt{\Omega_{m,0}a_\text{eq}}\ , \label{aR}\\
    \eta_\text{R}&=& 
    8\times10^9\,\mathrm{s}\,\left(\frac{2}{1+3w}\right)\left[\frac{g_*(T_\text{R})}{10.75}\right]^{-\frac{1}{6}}\left(\frac{1 \text{MeV}}{T_\text{R}}\right)\ ,  \label{etaR} 
\end{eqnarray}
where $g_\ast$ is an effective number of relativistic species as a function of the cosmic temperature $T_\text{R}$ corresponding to $\eta_\text{R}$, $H_0$ is the Hubble constant, $\Omega_\text{m,0}$ is the present-day energy-density fraction of dark matter, and $a_{\mathrm{eq}}$ denotes the scalar factor at the epoch of matter-radiation equality. 
The conformal Hubble parameter $\mathcal{H}=a'/a$ is 
\begin{eqnarray}
    \mathcal{H} = \left\{\begin{array}{ll}
\frac{2}{(1+3w)\eta} & \quad\quad  \mathrm{for}~~\eta < \eta_\text{R} \\
\frac{1}{\eta-\frac{1}{2}(1-3w)\eta_\text{R}} & \quad\quad  \mathrm{for}~~\eta \geq \eta_\text{R}
\end{array} \right. \label{H} ~,
\end{eqnarray}
where the prime denotes a derivative with respect to $\eta$.

The equation of motion for \acp{SIGW} is shown as \cite{Ananda:2006af,Baumann:2007zm,Mollerach:2003nq,Assadullahi:2009jc,Espinosa:2018eve,Kohri:2018awv}
\begin{eqnarray}
h_{i   j}'' + 2\mathcal{H}h_{i   j}' - \Delta h_{i   j}  =  - 4 \Lambda_{i   j}^{a   b} \mathcal{S}_{a   b} ~,\label{eom1}
\end{eqnarray}
where $h_{ij}$ is the strain of \acp{SIGW}, $\Lambda^{ab}_{ij}$ is the transverse-traceless operator, and $\mathcal{S}_{ab}$ denotes the source term consisting of nonlinear couplings between the linear cosmological perturbations $\psi$, namely, $\mathcal{S}_{ab}\sim \psi^2$. 
Analytic formulas of $\mathcal{S}_{ab}$ can be explicitly obtained via calculating the non-linear cosmological perturbations with \texttt{xpand} \cite{Pitrou:2013hga}.

The energy-density fraction spectrum of \acp{SIGW} for the sub-horizon modes ($k\eta \gg 1$) is given by \cite{Espinosa:2018eve,Kohri:2018awv} 
\begin{eqnarray}
  \Omega_{\rm{GW}} (k,\eta)    = \frac{1}{48} \left( \frac{k}{\mathcal{H}} \right)^2 \left[\mathcal{P}_h (k,\eta) + \bar{\mathcal{P}}_h (k,\eta)\right] \ , \label{eds}
\end{eqnarray}
where $k$ is a wavenunber, and $\mathcal{P}_h (k,\eta)$ and $\bar{\mathcal{P}}_h (k,\eta)$ stand for the two-point correlators of $h_{ij}$ and $h_{ij}'$, respectively, defined as 
\begin{eqnarray}
  \langle h_{i   j, \bm{k}} h_{i   j, \bm{\bar{k}}} \rangle  &=&  2 (2 \pi)^3 \delta ( \bm{k} + \bm{\bar{k}} ) \frac{2 \pi^2}{k^3} \mathcal{P}_h (k, \eta)~, \label{cor1} \\
  \langle {h'_{i   j, \bm{k}}} {h'_{i   j, \bm{\bar{k}}}} \rangle &=&  2 (2 \pi)^3 \delta ( \bm{k} + \bm{\bar{k}} ) \frac{2 \pi^2}{k^3} \bar{\mathcal{P}}_h (k, \eta) \ .
  \label{cor2}
\end{eqnarray}
In fact, Eq.~(\ref{cor2}) is redundant, since it can be derived from Eq.~(\ref{cor1}), but it lets us express Eq.~(\ref{eds}) in a simple form. 
Semi-analytic formulas for $h_{i j, \bm{k}}$ and ${h'_{i j, \bm{k}}}$ can be obtained explicitly. 
According to Eq.~(\ref{eom1}), the stochastic nature of $h_{i   j, \bm{k}}$ 
is attributed to that of $\Psi_{\bm{k}}$, i.e., the initial conditions of $\psi$.
The two-point correlator of $\Psi_{\bm{k}}$ is defined as 
\begin{eqnarray}
  \left\langle \Psi_{\bm{k}} \Psi_{\bm{\bar{k}}} \right\rangle & = & (2 \pi)^3 \delta \left( \bm{k} + \bm{\bar{k}} \right) \left[ \frac{3 (1 + w)}{5 + 3 w} \right]^2 \frac{2 \pi^2}{k^3} \mathcal{P}_{\zeta} (k) ~,
\end{eqnarray}
where $\mathcal{P}_{\zeta} (k)$ is a power spectrum of primordial curvature perturbations $\zeta$, and we have related $\psi$ to $\zeta$. 
For simplicity, we consider a monochromatic spectrum, i.e., $\mathcal{P}_\zeta=A k_\ast\delta(k-k_\ast)$, with $k_\ast$ being a reference wavenumber. 
Such a spectrum may be related to the formation of \acp{PBH} \cite{Hawking:1971ei,KHLOPOV1980383,Khlopov:1985jw}, and has been extensively studied in the literature (see reviews in Refs.~\cite{Carr:2020xqk,Sasaki:2018dmp} and references therein). 
It would be interesting to study a finite-width spectrum \cite{Papanikolaou:2022chm,Basilakos:2023xof}, which may change parts of our results. 
However, such a study is left to future works.

    {
    \centering
    \includegraphics[width=1\columnwidth]{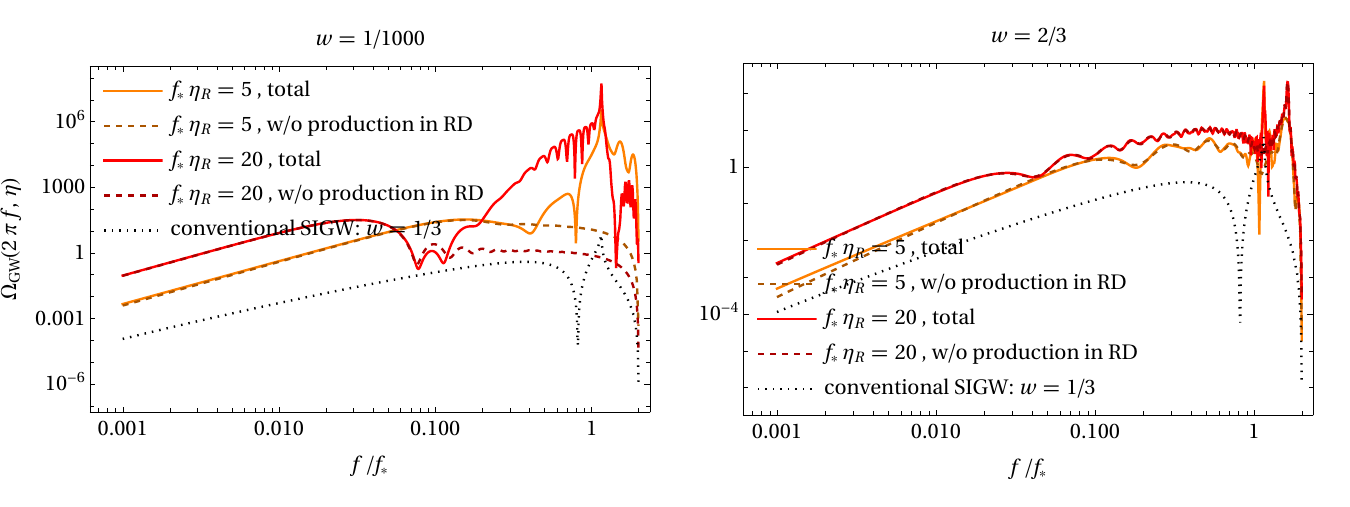}
    \figcaption{Energy-density fraction spectrum of SIGWs produced during the $w$-domination (dashed) versus the total spectrum (solid). For comparison, we show the conventional scenario of \acp{SIGW} (dotted).    \label{F2}}}

In Fig.~\ref{F2}, we show the energy-density fraction spectra of \acp{SIGW} for several different sets of model parameters. 
We take $k=2\pi f$ and $k_\ast=2\pi f_\ast$ throughout this work. 
For the small $w$, i.e., $w\rightarrow{0}$, the spectrum experiences an enhancement due to the sudden transition. 
In other words, when the early Universe transits from the almost early-matter domination to the radiation domination, the \acp{SIGW} produced during the radiation domination dominates the spectrum. 
This result is consistent with the enhancement mechanism proposed in Refs.~\cite{Inomata:2020tkl,Inomata:2019ivs}, where the authors considered a sudden transition from the early-matter domination to the radiation domination.

\section{Data analysis and results}\label{sec:method}

To compare the theoretical results with the observations, we should know the energy-density fraction spectrum of \acp{SIGW} in the present Universe, i.e., 
\begin{equation}
     \Omega_{\mathrm{SIGW}}(f) \simeq \Omega_{\mathrm{r},0} \times \Omega_{\mathrm{GW}}(k,\eta) \ ,\label{eq:ogw0}
\end{equation}
where $\Omega_{\mathrm{GW}}$ has been shown in Eq.~(\ref{eds}), and the physical energy-density fraction of radiations in the present-day Universe is $h^{2}\Omega_{\mathrm{r},0} \simeq 4.2\times10^{-5}$ with the dimensionless Hubble constant being $h=0.6766$, as measured by the Planck satellite \cite{Planck:2018vyg}.

Besides the above \ac{SIGW} component, the \ac{PTA} signal may be contributed by an astrophysical component arising from the binary \acp{SMBH}. 
The corresponding energy-density spectrum is parameterized as a power-law \cite{NANOGrav:2023hfp,EPTA:2023xxk}
\begin{equation}
    \Omega_{\rm BHB}(f) = \frac{2 \pi^2 f_{\mathrm{yr}}^2}{3 H_0^2}  A^2_{\mathrm{BHB}} \left(\frac{f}{f_{\mathrm{yr}}} \right)^{5-\gamma_{\rm BHB}}\ ,\label{eq:bhb}
\end{equation}
where $A_{\mathrm{BHB}}$ is the spectral amplitude, $\gamma_{\mathrm{BHB}}$ is the spectral index, and $f_{\mathrm{yr}}$ is a pivot frequency corresponding to one year.
Therefore, the total spectrum is expressed as a sum of Eq.~(\ref{eq:ogw0}) and Eq.~(\ref{eq:bhb}).

Using the same statistical tools used by the \ac{PTA} collaborations, we analyze the \ac{NG} 15-year data \cite{NANOGrav:2023gor} by using the publicly-available \texttt{PTArcade} \cite{andrea_mitridate_2023,Mitridate:2023oar}, which implements the Bayes parameter inferences for \ac{PTA} data by providing a wrapper of \texttt{ENTERPRISE} \cite{enterprise} and \texttt{Ceffyl} \cite{lamb2023rapid}. 
Here, the total spectrum would be fitted to the \ac{PTA} data during our data analysis.

We take the \ac{BBN} \cite{Cooke:2013cba} and \ac{CMB} \cite{Clarke:2020bil} constraints on the effective number of relativistic species into account, which can further pin down the parameter space of \acp{SIGW}. 
In terms of the integrated energy-density fraction, as defined by $\int_{k_{\mathrm{min}}}^{\infty}d\ln k \ h^{2}\Omega_{\mathrm{SIGW}}(k)$, the upper limits at 95\% confidence level are given as $1.3\times10^{-6}$ for \ac{BBN} \cite{Cooke:2013cba} and $2.9\times10^{-7}$ for \ac{CMB} \cite{Clarke:2020bil}. 
The lower bound of the integral is denoted as $k_{\rm min} = 2 \pi f_{\rm min}$, where $f_{\rm min}$ is $1.5\times10^{-11}\mathrm{Hz}$ for \ac{BBN} and $3\times10^{-17}$ for \ac{CMB} \cite{Maggiore:2018sht}. 
The \ac{BBN} and \ac{CMB} data have also been taken into account in studies of \acp{SIGW} \cite{Wang:2023sij} and phase-transition gravitational waves \cite{Bringmann:2023opz}. 
Here, we take each of them into account by assuming the integrated energy-density spectrum to follow a Gaussian distribution, whose variance is given by one-half of the aforementioned upper limit. 
Therefore, we adopt a modified version of \texttt{Ceffyl} \cite{lamb2023rapid} via adding two likelihoods to account for the \ac{BBN} and \ac{CMB} constraints, respectively.

The parameter space under investigation is spanned by $\log_{10}A$, $\log_{10}(f_\ast/\mathrm{Hz})$, $w$, $\log_{10}(f_\ast\eta_R)$, $\log_{10}A_{\mathrm{BHB}}$, and $\gamma_{\mathrm{BHB}}$. 
We adopt the uniform priors for the \ac{SIGW} parameters, i.e., $\log_{10} A\in \mathcal{U}[-8,1]$, $\log_{10}(f_\ast/\mathrm{Hz})\in\mathcal{U}[-9,-4.5]$, $w\in\mathcal{U}[0,1]$, and $\log_{10}(f_\ast \eta_\text{R}) \in \mathcal{U}[0,5]$. 
The priors for the last two parameters indicate a prior for $\eta_{\text{R}}$ within $[10^{9},10^{14}]$ seconds. 
According to Eq.~(\ref{etaR}), this prior promises $T_{\text{R}}$ to be higher than $\sim5$\,MeV, i.e., a temperature at which the $w$-domination should transition to the radiation domination, as suggested in Refs.~\cite{Sobotka:2022vrr,deSalas:2015glj,Hasegawa:2019jsa}. 
In addition, the priors for the \ac{SMBH} parameters, i.e., $\log_{10}A_{\mathrm{BHB}}$ and $\gamma_{\mathrm{BHB}}$, are the same as those of Ref.~\cite{NANOGrav:2023hvm}, which follow a bivariate Gaussian distribution, as wrapped in \texttt{PTArcade}.

{\centering
\includegraphics[width=1\columnwidth]{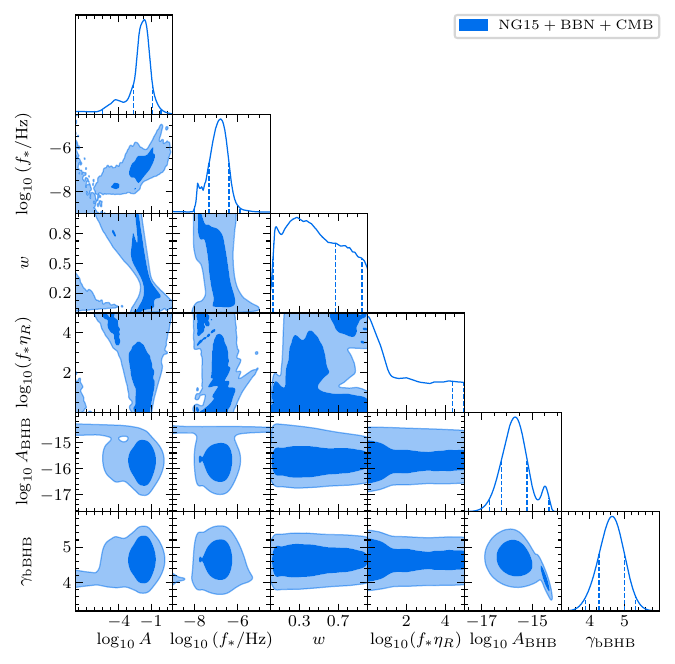}
\figcaption{Posteriors of six independent parameters.}\label{fig:posteriors1}}

{\tabcaption{Posterior and prior distributions for the parameters in the analysis. Here, $\mathcal{U}$ denotes the uniform distribution. We show the results at $95 \%$ confidence level.}
\centering
    \resizebox{ 1 \columnwidth}{!}{ 
    { \centering 
    \begin{tabular}{c | l r } 
        Parameter & Prior& Posterior\\
        \hline
        $\log_{10} A$ &  $\mathcal{U}[-8,1]$&$-1.92^{+1.50}_{-4.32}$\\
        $\log_{10} (f_{\star}/{\rm Hz})$& $\mathcal{U}[-9,-4.5]$&$-6.89^{+1.16}_{-1.03 }$\\
        $w$ & $\mathcal{U}[0,1]$&$0.44^{+0.52 }_{-0.40 }$\\
        $\log_{10} (f_{\ast}\eta_\text{R})$ & $\mathcal{U}[0,5]$&$1.74^{+3.09 }_{-1.68}$\\
        $\log_{10} A_{\mathrm{BHB}}$ & Same as Ref.~\cite{NANOGrav:2023hvm}& $-15.68^{+1.24 }_{-1.00}$\\
        $\gamma_{\mathrm{BHB}}$ & Same as Ref.~\cite{NANOGrav:2023hvm} &$4.63^{+0.69}_{-0.75}$
    \end{tabular}}
\label{Tab:Posteriors}
    }}

{\centering
\includegraphics[width=1\columnwidth]{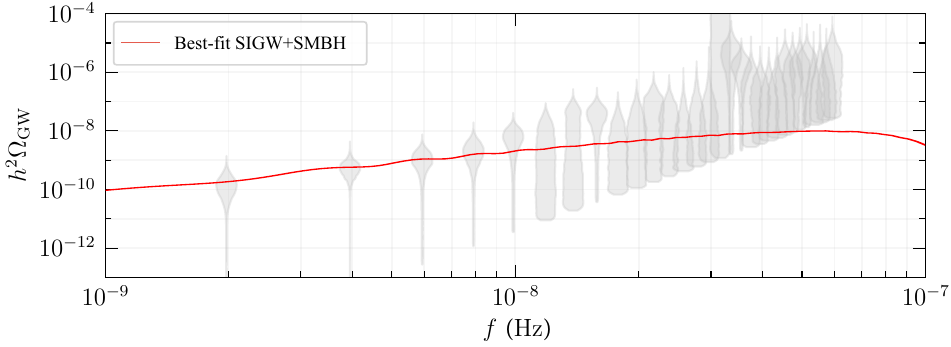}
\figcaption{Comparison between the best-fit spectrum and the NANOGrav signal.}\label{fig:compare}}

We depict the posteriors of six independent parameters in
Fig.~\ref{fig:posteriors1}. 
Compared with Ref.~\cite{NANOGrav:2023hvm}, which analyzed the \ac{NG} data only, we combine the \ac{NG} data with the \ac{BBN} and \ac{CMB} data. 
Hence, the inferred posteriors in our work are changed significantly. 
In particular, we find that at 95\% confidence level, $A$ is bounded from the upper, but not from the lower, i.e., $A\leq0.39$, and $f_{\ast}$ is within $1.2\times10^{-8}-1.9\times10^{-6}\,$Hz. 
In contrast, we get looser constraints on $w$ and $f_\ast\eta_R$. 
We find that most of the considered \ac{EoS} fit the joint dataset just as well as $w=1/3$ within 95\% confidence interval. 
In addition, our constraints on $A_{\mathrm{BHB}}$ and $\gamma_{\mathrm{BHB}}$ are compatible with those of Ref.~\cite{NANOGrav:2023hvm}. 
We summarize the information of priors and posteriors in Tab.~\ref{Tab:Posteriors}. 
Furthermore, we compare the best-fit spectrum with the observed signal in Fig.~\ref{fig:compare}.

    {
    \centering
    \includegraphics[width=1\columnwidth]{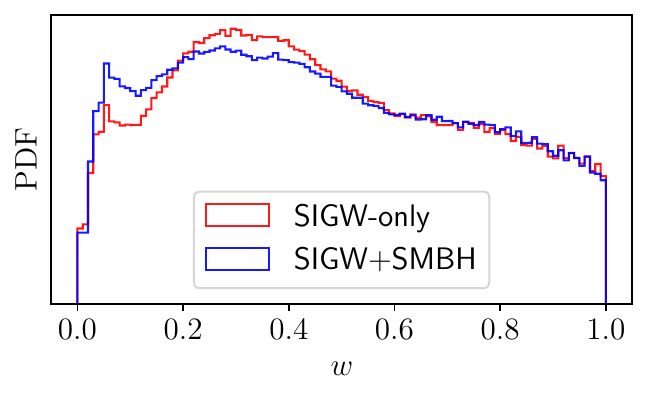}
    \figcaption{ Comparison of the probability distribution functions of the EoS parameter $w$ between the SIGW-only model (red) and the SIGW+SMBH model (blue).}    \label{F3}
    }

Following the same approach mentioned above, we perform an additional analysis of the \ac{SIGW}-only model, meaning that we consider the \ac{SIGW} parameters only and disregard the \ac{SMBH} parameters. 
In Fig.~\ref{F3}, we compare the inferred probability distribution functions of $w$ between the \ac{SIGW}-only model (red curve) and the combined \ac{SIGW}+\ac{SMBH} model (blue curve). 
We find that they are compatible with each other.

\section{PBH production \label{III}}

The \ac{PBH} formation during the radiation domination results in the overproduction of \acp{PBH} \cite{NANOGrav:2023hvm,EPTA:2023xxk}. 
In contrast, we expect that \acp{PBH} were produced during an epoch instead of the radiation domination. 
The \ac{PBH} abundance can be changed to some extent, possibly leading to avoidance of the overproduction issue.

When the $k$ modes reentered into the Hubble horizon, i.e., $k\simeq\mathcal{H}=2/[(1+3w)\eta]$, and collapsed to form \acp{PBH}, the \ac{PBH} fraction of dark matter is given by \cite{Sasaki:2018dmp}
\begin{eqnarray}
    f_\text{pbh}(k)&=&\frac{a^3}{\Omega_\text{CDM}}\left(\frac{{H}}{{H}_0}\right)^2\gamma\beta(k)  \\ 
     &=& \frac{\Omega_{m,0}}{\Omega_\text{CDM}}\left[\frac{a_\text{eq}}{a_\text{R}}\right]\left(\frac{\eta}{\eta_\text{R}}\right)^{-\frac{6w}{1+3w}}\gamma\beta(k) \bigg|_{\eta=\frac{2}{(1+3w)k}}~, \nonumber
\end{eqnarray}
where $H$ is the Hubble parameter, and we have $\gamma=0.356$ \cite{NANOGrav:2023hvm}, $\Omega_{m,0}=0.3089$, and $\Omega_\text{CDM} = 0.2589$ \cite{Planck:2015fie}. 
The probability of \ac{PBH} production is defined as \cite{Press:1973iz}
\begin{eqnarray}
    \beta(k)&=& \int_{\delta_c}^{\infty}\frac{1}{\sqrt{2\pi}\sigma{(k)}}\exp{\left(-\frac{\delta^2}{2\sigma(k)}\right)}\mathrm{d}\delta   \nonumber\\
    &=& \frac{1}{2}\mathrm{erfc}\left( \frac{\delta_c}{\sqrt{2}\sigma(k)}\right)~,
\end{eqnarray}
where the critical overdensity for gravitational collapse is $\delta_c= [{3(1+w)}/{(5+3w)}]\sin^2\left[ {(\pi\sqrt{w})}/{(1+3w)}\right]$ \cite{Harada:2016mhb}.
The coarse-grained overdensity is defined as 
\begin{eqnarray}
        \sigma^2(k)= \int^\infty_{0} \mathrm{d}\ln{q}W^2(q,R) \mathcal{P}_\delta(q,\eta)\Big|_{R=\eta=\frac{2}{(1+3w)k}}~,
\end{eqnarray}
where $W(k,R)=\exp[-(k R)^2/2]$ is a window function, and the power spectrum of overdensity is related to that of primordial curvature perturbations $\zeta$, i.e., 
$\mathcal{P}_\delta=\left[{2(1+w)}/{(5+3w)}\right]^2(k/\mathcal{H})^4 T^2(k\eta)\mathcal{P}_{\zeta}(k)$ with $T(k\eta)$ being the transfer function of $\psi$. 
One should note that the transfer function for the case of $w\neq1/3$ is different from that for the radiation domination.

To define the \ac{PBH} abundance, we should derive a relation between the \ac{PBH} mass $M_{\text{PBH}}$ and $k$ via evaluating $k/k_\ast=\mathcal{H}/\mathcal{H}_\ast$ by making use of $M_\text{PBH}=\gamma M_H= \gamma (4\pi/3)\rho {{H}}^{-3}$ and Eq.~(\ref{a}).
Such a relation is 
\begin{eqnarray}
    \frac{k}{k_\ast}=\left(\frac{M_\text{PBH}}{M_{\text{PBH},\ast}}\right)^{-\frac{1+3w}{3(1+w)}}~,
\end{eqnarray}
where $M_{\text{PBH},\ast}$ corresponds to $k_\ast$. 
Finally, the PBH abundance is defined as \cite{Ando:2018qdb, NANOGrav:2023hvm}
\begin{eqnarray}
    f_\text{PBH} &=& \int_{0}^{M_{\text{PBH,R}}} f_\text{pbh}(k(M_\text{PBH}))\mathrm{d}\ln{M_\text{PBH}}\nonumber \\
    &=&\frac{3(1+w)}{1+3w}\int_{\frac{2}{(1+3w)\eta_\text{R}}}^\infty f_\text{pbh}(k)\mathrm{d}\ln{k}~,
\end{eqnarray}
where $M_\text{PBH,R}$ corresponds to the $k$ modes reentering into the Hubble horizon at $\eta_\text{R}$.

{\centering
\includegraphics[width=1 \columnwidth]{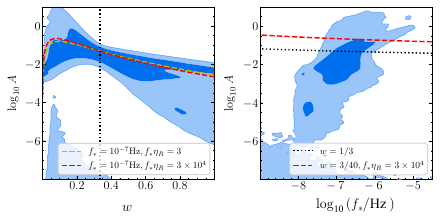}
  \figcaption{ The PBH overproduction versus the SIGW scenario with a sudden transition from the epoch of $w\neq1/3$ to the radiation domination. We denote $f_{\mathrm{PBH}}=1$ as dashed curves, above which PBHs are over-produced. For comparison, posteriors are shown as shaded regions. In the left panel, $w=1/3$ is shown in the dotted line. In the right panel, $f_{\mathrm{PBH}}=1$ for $w=1/3$ is also shown in the dotted curve.  \label{fig:main}
}}
  

In Fig.~\ref{fig:main}, we show to what extent we could resolve the issue of \ac{PBH} overproduction via comparing $f_{\mathrm{PBH}}=1$, as predicted by our scenarios, with the inferred posteriors. 
As suggested in Ref.~\cite{NANOGrav:2023hvm}, where only the \ac{PTA} data was analyzed, the issue of \ac{PBH} overproduction is inevitable in the case of $w=1/3$. 
Though being changed significantly, as shown in Sec.~{\ref{sec:method}}, the posteriors associating with the combination of \ac{PTA}, \ac{BBN}, and \ac{CMB} still can not avoid the \ac{PBH} overproduction. 
In this sense, we pay attention to scenarios of the varying \ac{EoS} in the early Universe. 
It was surprising to find that the scenarios of $w\sim\mathcal{O}(10^{-2})$, which are allowed by our joint dataset, can alleviate the overproduction issue.

\section{Conclusions and discussion}\label{sec:conclusion}

In this work, we have studied the influence of the \ac{EoS} of the early Universe on the production of \acp{SIGW}, and further interpreted the recent \ac{PTA} data releases as well as the overproduction issue of \acp{PBH}. 
Via generalizing the scenario of Refs.~\cite{Kohri:2018awv,Inomata:2020tkl,Inomata:2019ivs} and following the method of Refs.~\cite{Hajkarim:2019nbx,Domenech:2019quo,Domenech:2021ztg}, we computed the energy-density fraction spectrum of \acp{SIGW} for a sudden transition of the early Universe from an arbitrary \acp{EoS} $w\in[0,1]$ to the radiation domination $w=1/3$. 
We showed that the \ac{SIGW} spectrum exhibits a notable enhancement as the parameter $w$ approaches zero, as shown in Fig.~\ref{F2}. 
To get precise constraints on the model parameters, we analyzed the data combination incorporating the \ac{NG} 15-year \ac{PTA} data with the \ac{BBN} and \ac{CMB} data. 
To our knowledge, it is the first time to perform Bayes data analysis for the considered topic. 
It is also the first time to combine the \ac{PTA} data with the \ac{BBN} and \ac{CMB} data for the Bayes data analysis. 
The posteriors of model parameters have been shown in Tab.~\ref{Tab:Posteriors}. 
In addition, studying the \ac{PBH} production in the $w$-dominated epoch, we found that the \ac{PBH}-overproduction issue can be alleviated to some extent, as demonstrated by Fig.~\ref{fig:main}.

In fact, the production of \acp{SIGW} in the early Universe with an arbitrary \ac{EoS} was studied in Refs.~\cite{Hajkarim:2019nbx,Domenech:2019quo,Domenech:2020kqm,Domenech:2021ztg}. 
In these works, the gravitational waves were produced during the $w$-domination, without further production during the subsequent radiation domination. 
However, it was shown that for \acp{SIGW} produced during both the early-matter domination and subsequent radiation domination, the \ac{SIGW} spectrum could be significantly enhanced due to a sudden transition of \ac{EoS} \cite{Inomata:2020tkl,Inomata:2019ivs}. 
We have found the similar enhancement mechanism for our scenario.

We have proposed that the \ac{SIGW} spectrum is tightly constrained by the \ac{PTA} data and the total energy-density fraction by the \ac{BBN}, and \ac{CMB} data. 
In Tab.~\ref{Tab:Posteriors}, we obtained the allowed parameter region, e.g., the spectral amplitude $A\leq0.39$, spectral peak frequency $f_\ast\sim10^{-8}-10^{-6}$ Hz, \ac{EoS} $w=0.44_{-0.40}^{+0.52}$, etc. 
Compared with those of Ref.~\cite{NANOGrav:2023hvm}, the posteriors have been significantly changed due to combining the \ac{PTA} data with the \ac{BBN} and \ac{CMB} data and considering different scenarios of \acp{SIGW}. 
We revealed the importance of \ac{BBN} and \ac{CMB} for data analysis. 
Our results can be further tested with future observations. 
Other works relevant to the background \ac{EoS} can be found in Ref.~\cite{Oikonomou:2023qfz}.

Through exploring the production of \acp{PBH} during the epoch of $w$-domination, we have found that our scenario of $w\sim\mathcal{O}(10^{-2})$ can suppress the abundance of \acp{PBH}, leading to alleviation of the overproduction issue \cite{NANOGrav:2023hvm,EPTA:2023xxk}. 
One should note that we derived the systematic formulas for the \ac{PBH} abundance, implying a consistent study of \ac{PBH} production. 
In addition, the speed of transition from the $w$-domination to the radiation domination is another concern to investigate. 
Introducing a finite speed of transition into our model may change our results \cite{Inomata:2019zqy,Inomata:2020lmk}. 
However, it is extremely complicated to take the effect of transition speed into account, following the existing research approach. 
On the other hand, to our knowledge, there are not any works along this research direction until now. 
We think that such a study warrants for another paper, which is left to future works.

\

\acknowledgments{
		We acknowledge Zucheng Chen, Shi Pi and Yan-Heng Yu for helpful discussion. Q.H.Z is supported by the National Natural Science Foundation of China (Grant NO. 12305073). Z.C.Z. is supported by the National Key Research and Development Program of China Grant No. 2021YFC2203001 and the National Natural Science Foundation of China (Grant NO. 12005016). S.W. is partially supported by the National Natural Science Foundation of China (Grant No. 12175243), the National Key R\&D Program of China No. 2023YFC2206403, the Science Research Grants from the China Manned Space Project with No. CMS-CSST-2021-B01, and the Key Research Program of the Chinese Academy of Sciences (Grant No. XDPB15). X.Z. is supported by the National SKA Program of China (Grants Nos. 2022SKA0110200 and 2022SKA0110203) and the National Natural Science Foundation of China (Grants Nos. 11975072, 11835009, and 11805031). This work is supported by High-performance Computing Platform of China Agricultural University.}

\end{multicols}

\ 

\ 

\appendix


\section{Appendix: SIGWs due to a sudden transition of the early Universe \label{appA}}

In conformal Newtonian gauge, the perturbed spatially-flat \ac{FRW} metric is 
\begin{eqnarray}
{\rm d} s^2 & = & a^2  \left\{ - \left(1 + 2 \phi\right)  {\rm d} t^2 + \left[ \left(1 - 2 \psi\right) \delta_{i   j} + \frac{1}{2} h_{i   j} \right] {\rm d} x^i  {\rm d} x^j \right\}\ ,
\end{eqnarray}
where $\eta$ is a conformal time, $a $ is a scale factor of the Universe, both $\phi$ and $\psi$ are the linear cosmological scalar perturbations, and $h_{ij}$ denotes the second-order tensor perturbations, i.e., \acp{SIGW}. 
In the absence of anisotropic stress, we have $\psi=\phi$, exactly. 
Here, the evolution of $\psi$ is determined by the master equation,
\begin{eqnarray}
\psi'' + 3 (1 + w) \mathcal{H} \psi' - w \Delta \psi & = & 0 \ ,  \label{eom2}
\end{eqnarray}
where $'$ denotes the derivative with respect to $\eta$, and we have adopted an adiabatic sound speed $c_s^2 = w$ with $0\leq w \leq 1$ for a phenomenological study. 
By solving Eq.~(\ref{eom2}), we obtain the Fourier modes of $\psi$ in the $w$-domination and radiation domination, respectively, namely, 
\begin{eqnarray}
\psi_{\bm{k}}^{({w})} & = &  _0 F_1 \left( \frac{7 + 9 w}{2 (1 + 3) w} ; - \frac{1}{4} w   k^2 \eta^2 \right) \Psi_{\bm{k}}\ \label{A8},
  \\
\psi_{\bm{k}} & = &   \left[ \frac{- \sqrt{3} k \eta_w \cos \left( {k \eta_w}/{\sqrt{3}} \right) + 3 \sin \left( {k \eta_w}/{\sqrt{3}} \right)}{(k \eta_w)^3} \right] \psi_{\bm{k}}^{(\text{A})}  + \left[ \frac{- 3 \cos \left( {k \eta_w}/{\sqrt{3}} \right) - \sqrt{3} k \eta_w \sin \left( {k \eta_w}/{\sqrt{3}} \right)}{(k \eta_w)^3} \right] \psi^{(\text{B})}_{\bm{k}}\ , \nonumber\\
&&
\end{eqnarray}
where the superscripts ($w$) denote the $w$-domination, we introduce $\eta_{w}= \eta-(1-3w)\eta_\text{R}/2$, $\Psi_{\bm{k}}$ stands for a random variable related to the primordial curvature perturbations, and $_0 F_1$ is the confluent hypergeometric function. 
According to the continuity and differentiability of $\psi_{\bm k}$, we can determine $\psi_{\bm{k}}^{(\text{A})}$ and $\psi_{\bm{k}}^{(\text{B})}$ as follows 
\begin{eqnarray}
  \psi_{\bm{k}}^{(\text{A})} & = & \left[ \frac{(y^2 - 9)}{\sqrt{3}}
  \cos \left( \frac{y}{\sqrt{3}} \right) + 3 y  \sin \left(
  \frac{y}{\sqrt{3}} \right) \right] \left. \psi_{\bm{k}}^{({w})}
  \right|_{\eta_\text{R}} + y \left[ \sqrt{3}  \cos \left( \frac{y}{\sqrt{3}}
  \right) + y  \sin \left( \frac{y}{\sqrt{3}} \right) \right] \left.
  k^{-1}{\psi_{\bm{k}}^{({w})}}' \right|_{\eta_\text{R}} \ ,\nonumber\\
  &&\\
  \psi_{\bm{k}}^{(\text{B})} & = & \left[ - \sqrt{3 } y
   \cos \left( \frac{y}{\sqrt{3}} \right) + (y^2 - 9) \sin \left(
  \frac{y}{\sqrt{3}} \right) \right] \left. \psi_{\bm{k}}^{({w})}
  \right|_{\eta_\text{R}} - y \left[ y  \cos \left( \frac{y}{\sqrt{3}}
  \right) - \sqrt{3} \sin \left( \frac{y}{\sqrt{3}} \right) \right] \left.
  k^{-1}{\psi_{\bm{k}}^{({w})}}' \right|_{\eta_\text{R}} \ ,\nonumber\\
  &&
\end{eqnarray}
where we have introduced $y= k\eta_w$.

The equation of motion for \acp{SIGW} has been presented in Eq.~(5) in main text,
where the source term $\mathcal{S}_{a   b}$ consisting of the linear perturbations is given by 
\begin{eqnarray}
\mathcal{S}_{a   b} & = & \frac{2 (5 + 3 w)}{3 (1 + w)} \partial_a \psi \partial_b \psi + \frac{4}{3 (1 + w) \mathcal{H}} (\partial_a \psi \partial_b \psi' + \partial_a \psi' \partial_b \psi') + \frac{4}{3 (1 + w) \mathcal{H}^2} \partial_a \psi' \partial_b \psi' \ .  \label{src}
\end{eqnarray}
Therefore, in the Fourier space, we obtain the subhorizon modes of $h_{i   j,\bm k}$ based on Eq.~(5) in the main text, namely, 
\begin{eqnarray}
  h_{i j, \bm{k}} 
  &=&\frac{\sin (k \eta)}{k \eta_w} h_{i  j,
  \bm{k}}^{(\text{A})} + \frac{\cos (k \eta)}{k \eta_w} h^{(
  \text{B})}_{i  j, \bm{k}} + \frac{\sin (k \eta)}{k \eta_w}
  \int_{\eta_\text{R}}^{\eta} {\rm d} \bar{\eta} \left\{  \cos (k \bar{\eta}) \Lambda^{a b}_{i  j} (\hat{\bm k}) \mathcal{S}_{a  b,\bm{k}} (\bar{\eta})\big|_{w=1/3} \left[ \bar{\eta} -  \frac{1}{2} (1 - 3 w) \eta_\text{R} \right] \right\} \nonumber\\
  &  & - \frac{\cos (k \eta)}{k \eta_w} \int_{\eta_\text{R}}^{\eta} {\rm d}\bar{\eta} \left[ \sin  (k \bar{\eta}) \Lambda^{a  b}_{i  j}(\hat{\bm k}) \mathcal{S}_{a  b, \bm{k}} (\bar{\eta})\big|_{w=1/3} \left( \bar{\eta} - \frac{1}{2} (1 - 3 w) \eta_\text{R} \right)
  \right] ~, \label{h2}
\end{eqnarray}
where $\mathcal{S}_{ab,\bm{k}}$ stands for the Fourier mode of $\mathcal{S}_{ab}$ with the wavevector $\bm{k}$. 
Due to the continuity and differentiability of $h_{ij,\bm k}$, the time-independent quantities $h_{i   j, \bm{k}}^{(\text{A})}$ and $h^{(\text{B})}_{i   j, \bm{k}}$ can be determined by \acp{SIGW} produced during the $w$-domination (denoted as $h_{i j, \bm{k}}^{(w)})$, namely,
\begin{eqnarray}
  h^{(\text{A})}_{i   j, \bm{k}} & = &  \eta_w \cos (k \eta_\text{R}) \left. {h_{i   j, \bm{k}}^{({w})}}' \right|_{\eta = \eta_\text{R}} + \left[\cos (k \eta_\text{R}) + k \eta_w \sin (k \eta_\text{R})\right] \left. h_{i   j, \bm{k}}^{({w})} \right|_{\eta = \eta_\text{R}} \ , \\
  h^{(\text{B})}_{i   j, \bm{k}} & = & \left[ k \eta_w \cos (k \eta_\text{R}) -\sin(k\eta_\text{R}) \right]  \left. h_{i   j, \bm{k}}^{({w})} \right|_{\eta = \eta_\text{R}} -\eta_w  \sin (k \eta_\text{R}) \left. {h_{i   j, \bm{k}}^{({w})}}' \right|_{\eta = \eta_\text{R}}   \ . \label{hrd} 
\end{eqnarray}
where we define  
\begin{eqnarray}
      h_{i j, \bm{k}}^{(w)} &=& (k {\eta})^{\alpha} Y_{\alpha} (k \eta) \int_0^{\eta} \frac{{\rm d} \bar{\eta}}{k} \left[ \frac{\pi}{2} (k \bar{\eta})^{1 - \alpha} J_{\alpha} (k \bar{\eta}) \Lambda^{a b}_{i j} (\hat{\bm k}) \mathcal{S}_{a b, \bm{k}} \right]  - (k \eta)^{\alpha} J_{\alpha} (k \eta) \int_0^{\eta} \frac{{\rm d} \bar{\eta}}{k} \left[ (k \bar{\eta})^{1 - \alpha} Y_{\alpha} (k \bar{\eta}) \Lambda^{a b}_{i j} (\hat{\bm k}) \mathcal{S}_{a b, \bm{k}} \right]\ , \nonumber\\
  \label{h1} 
\end{eqnarray}
and we introduce $\alpha = - 3 (1 - w) / [2 (1 + 3 w)]$, $Y_{\alpha}(x)$ is the Bessel function of second kind, and $J_{\alpha}(x)$ is the Bessel function of first kind.


\

\begin{multicols}{2}
\bibliography{pgw}

\begin{thebibliography}{89}%
\makeatletter
\providecommand \@ifxundefined [1]{%
 \@ifx{#1\undefined}
}%
\providecommand \@ifnum [1]{%
 \ifnum #1\expandafter \@firstoftwo
 \else \expandafter \@secondoftwo
 \fi
}%
\providecommand \@ifx [1]{%
 \ifx #1\expandafter \@firstoftwo
 \else \expandafter \@secondoftwo
 \fi
}%
\providecommand \natexlab [1]{#1}%
\providecommand \enquote  [1]{``#1''}%
\providecommand \bibnamefont  [1]{#1}%
\providecommand \bibfnamefont [1]{#1}%
\providecommand \citenamefont [1]{#1}%
\providecommand \href@noop [0]{\@secondoftwo}%
\providecommand \href [0]{\begingroup \@sanitize@url \@href}%
\providecommand \@href[1]{\@@startlink{#1}\@@href}%
\providecommand \@@href[1]{\endgroup#1\@@endlink}%
\providecommand \@sanitize@url [0]{\catcode `\\12\catcode `\$12\catcode
  `\&12\catcode `\#12\catcode `\^12\catcode `\_12\catcode `\%12\relax}%
\providecommand \@@startlink[1]{}%
\providecommand \@@endlink[0]{}%
\providecommand \url  [0]{\begingroup\@sanitize@url \@url }%
\providecommand \@url [1]{\endgroup\@href {#1}{\urlprefix }}%
\providecommand \urlprefix  [0]{URL }%
\providecommand \Eprint [0]{\href }%
\providecommand \doibase [0]{http://dx.doi.org/}%
\providecommand \selectlanguage [0]{\@gobble}%
\providecommand \bibinfo  [0]{\@secondoftwo}%
\providecommand \bibfield  [0]{\@secondoftwo}%
\providecommand \translation [1]{[#1]}%
\providecommand \BibitemOpen [0]{}%
\providecommand \bibitemStop [0]{}%
\providecommand \bibitemNoStop [0]{.\EOS\space}%
\providecommand \EOS [0]{\spacefactor3000\relax}%
\providecommand \BibitemShut  [1]{\csname bibitem#1\endcsname}%
\let\auto@bib@innerbib\@empty
\bibitem [{\citenamefont {Xu}\ \emph {et~al.}(2023)\citenamefont {Xu} \emph
  {et~al.}}]{Xu:2023wog}%
  \BibitemOpen
  \bibfield  {author} {\bibinfo {author} {\bibfnamefont {H.}~\bibnamefont {Xu}}
  \emph {et~al.},\ }\href {\doibase 10.1088/1674-4527/acdfa5} {\bibfield
  {journal} {\bibinfo  {journal} {Res. Astron. Astrophys.}\ }\textbf {\bibinfo
  {volume} {23}},\ \bibinfo {pages} {075024} (\bibinfo {year} {2023})},\
  \Eprint {http://arxiv.org/abs/2306.16216} {arXiv:2306.16216 [astro-ph.HE]}
  \BibitemShut {NoStop}%
\bibitem [{\citenamefont {Antoniadis}\ \emph {et~al.}(2023)\citenamefont
  {Antoniadis} \emph {et~al.}}]{EPTA:2023fyk}%
  \BibitemOpen
  \bibfield  {author} {\bibinfo {author} {\bibfnamefont {J.}~\bibnamefont
  {Antoniadis}} \emph {et~al.} (\bibinfo {collaboration} {EPTA, InPTA:}),\
  }\href {\doibase 10.1051/0004-6361/202346844} {\bibfield  {journal} {\bibinfo
   {journal} {Astron. Astrophys.}\ }\textbf {\bibinfo {volume} {678}},\
  \bibinfo {pages} {A50} (\bibinfo {year} {2023})},\ \Eprint
  {http://arxiv.org/abs/2306.16214} {arXiv:2306.16214 [astro-ph.HE]}
  \BibitemShut {NoStop}%
\bibitem [{\citenamefont {Agazie}\ \emph
  {et~al.}(2023{\natexlab{a}})\citenamefont {Agazie} \emph
  {et~al.}}]{NANOGrav:2023gor}%
  \BibitemOpen
  \bibfield  {author} {\bibinfo {author} {\bibfnamefont {G.}~\bibnamefont
  {Agazie}} \emph {et~al.} (\bibinfo {collaboration} {NANOGrav}),\ }\href
  {\doibase 10.3847/2041-8213/acdac6} {\bibfield  {journal} {\bibinfo
  {journal} {Astrophys. J. Lett.}\ }\textbf {\bibinfo {volume} {951}} (\bibinfo
  {year} {2023}{\natexlab{a}}),\ 10.3847/2041-8213/acdac6},\ \Eprint
  {http://arxiv.org/abs/2306.16213} {arXiv:2306.16213 [astro-ph.HE]}
  \BibitemShut {NoStop}%
\bibitem [{\citenamefont {Reardon}\ \emph {et~al.}(2023)\citenamefont {Reardon}
  \emph {et~al.}}]{Reardon:2023gzh}%
  \BibitemOpen
  \bibfield  {author} {\bibinfo {author} {\bibfnamefont {D.~J.}\ \bibnamefont
  {Reardon}} \emph {et~al.},\ }\href {\doibase 10.3847/2041-8213/acdd02}
  {\bibfield  {journal} {\bibinfo  {journal} {Astrophys. J. Lett.}\ }\textbf
  {\bibinfo {volume} {951}} (\bibinfo {year} {2023}),\
  10.3847/2041-8213/acdd02},\ \Eprint {http://arxiv.org/abs/2306.16215}
  {arXiv:2306.16215 [astro-ph.HE]} \BibitemShut {NoStop}%
\bibitem [{\citenamefont {Franciolini}\ \emph {et~al.}(2023)\citenamefont
  {Franciolini}, \citenamefont {Iovino}, \citenamefont {Vaskonen},\ and\
  \citenamefont {Veermae}}]{Franciolini:2023pbf}%
  \BibitemOpen
  \bibfield  {author} {\bibinfo {author} {\bibfnamefont {G.}~\bibnamefont
  {Franciolini}}, \bibinfo {author} {\bibfnamefont {A.}~\bibnamefont {Iovino},
  \bibfnamefont {Junior.}}, \bibinfo {author} {\bibfnamefont {V.}~\bibnamefont
  {Vaskonen}}, \ and\ \bibinfo {author} {\bibfnamefont {H.}~\bibnamefont
  {Veermae}},\ }\href {\doibase 10.1103/PhysRevLett.131.201401} {\bibfield
  {journal} {\bibinfo  {journal} {Phys. Rev. Lett.}\ }\textbf {\bibinfo
  {volume} {131}},\ \bibinfo {pages} {201401} (\bibinfo {year} {2023})},\
  \Eprint {http://arxiv.org/abs/2306.17149} {arXiv:2306.17149 [astro-ph.CO]}
  \BibitemShut {NoStop}%
\bibitem [{\citenamefont {Inomata}\ \emph {et~al.}(2024)\citenamefont
  {Inomata}, \citenamefont {Kohri},\ and\ \citenamefont
  {Terada}}]{Inomata:2023zup}%
  \BibitemOpen
  \bibfield  {author} {\bibinfo {author} {\bibfnamefont {K.}~\bibnamefont
  {Inomata}}, \bibinfo {author} {\bibfnamefont {K.}~\bibnamefont {Kohri}}, \
  and\ \bibinfo {author} {\bibfnamefont {T.}~\bibnamefont {Terada}},\ }\href
  {\doibase 10.1103/PhysRevD.109.063506} {\bibfield  {journal} {\bibinfo
  {journal} {Phys. Rev. D}\ }\textbf {\bibinfo {volume} {109}},\ \bibinfo
  {pages} {063506} (\bibinfo {year} {2024})},\ \Eprint
  {http://arxiv.org/abs/2306.17834} {arXiv:2306.17834 [astro-ph.CO]}
  \BibitemShut {NoStop}%
\bibitem [{\citenamefont {Cai}\ \emph {et~al.}(2023)\citenamefont {Cai},
  \citenamefont {He}, \citenamefont {Ma}, \citenamefont {Yan},\ and\
  \citenamefont {Yuan}}]{Cai:2023dls}%
  \BibitemOpen
  \bibfield  {author} {\bibinfo {author} {\bibfnamefont {Y.-F.}\ \bibnamefont
  {Cai}}, \bibinfo {author} {\bibfnamefont {X.-C.}\ \bibnamefont {He}},
  \bibinfo {author} {\bibfnamefont {X.-H.}\ \bibnamefont {Ma}}, \bibinfo
  {author} {\bibfnamefont {S.-F.}\ \bibnamefont {Yan}}, \ and\ \bibinfo
  {author} {\bibfnamefont {G.-W.}\ \bibnamefont {Yuan}},\ }\href {\doibase
  10.1016/j.scib.2023.10.027} {\bibfield  {journal} {\bibinfo  {journal} {Sci.
  Bull.}\ }\textbf {\bibinfo {volume} {68}},\ \bibinfo {pages} {2929} (\bibinfo
  {year} {2023})},\ \Eprint {http://arxiv.org/abs/2306.17822} {arXiv:2306.17822
  [gr-qc]} \BibitemShut {NoStop}%
\bibitem [{\citenamefont {Wang}\ \emph
  {et~al.}(2024{\natexlab{a}})\citenamefont {Wang}, \citenamefont {Zhao},
  \citenamefont {Li},\ and\ \citenamefont {Zhu}}]{Wang:2023ost}%
  \BibitemOpen
  \bibfield  {author} {\bibinfo {author} {\bibfnamefont {S.}~\bibnamefont
  {Wang}}, \bibinfo {author} {\bibfnamefont {Z.-C.}\ \bibnamefont {Zhao}},
  \bibinfo {author} {\bibfnamefont {J.-P.}\ \bibnamefont {Li}}, \ and\ \bibinfo
  {author} {\bibfnamefont {Q.-H.}\ \bibnamefont {Zhu}},\ }\href {\doibase
  10.1103/PhysRevResearch.6.L012060} {\bibfield  {journal} {\bibinfo  {journal}
  {Phys. Rev. Res.}\ }\textbf {\bibinfo {volume} {6}},\ \bibinfo {pages}
  {L012060} (\bibinfo {year} {2024}{\natexlab{a}})},\ \Eprint
  {http://arxiv.org/abs/2307.00572} {arXiv:2307.00572 [astro-ph.CO]}
  \BibitemShut {NoStop}%
\bibitem [{\citenamefont {Liu}\ \emph {et~al.}(2024)\citenamefont {Liu},
  \citenamefont {Chen},\ and\ \citenamefont {Huang}}]{Liu:2023ymk}%
  \BibitemOpen
  \bibfield  {author} {\bibinfo {author} {\bibfnamefont {L.}~\bibnamefont
  {Liu}}, \bibinfo {author} {\bibfnamefont {Z.-C.}\ \bibnamefont {Chen}}, \
  and\ \bibinfo {author} {\bibfnamefont {Q.-G.}\ \bibnamefont {Huang}},\ }\href
  {\doibase 10.1103/PhysRevD.109.L061301} {\bibfield  {journal} {\bibinfo
  {journal} {Phys. Rev. D}\ }\textbf {\bibinfo {volume} {109}},\ \bibinfo
  {pages} {L061301} (\bibinfo {year} {2024})},\ \Eprint
  {http://arxiv.org/abs/2307.01102} {arXiv:2307.01102 [astro-ph.CO]}
  \BibitemShut {NoStop}%
\bibitem [{\citenamefont {Abe}\ and\ \citenamefont {Tada}(2023)}]{Abe:2023yrw}%
  \BibitemOpen
  \bibfield  {author} {\bibinfo {author} {\bibfnamefont {K.~T.}\ \bibnamefont
  {Abe}}\ and\ \bibinfo {author} {\bibfnamefont {Y.}~\bibnamefont {Tada}},\
  }\href {\doibase 10.1103/PhysRevD.108.L101304} {\bibfield  {journal}
  {\bibinfo  {journal} {Phys. Rev. D}\ }\textbf {\bibinfo {volume} {108}},\
  \bibinfo {pages} {L101304} (\bibinfo {year} {2023})},\ \Eprint
  {http://arxiv.org/abs/2307.01653} {arXiv:2307.01653 [astro-ph.CO]}
  \BibitemShut {NoStop}%
\bibitem [{\citenamefont {Ebadi}\ \emph {et~al.}(2024)\citenamefont {Ebadi},
  \citenamefont {Kumar}, \citenamefont {McCune}, \citenamefont {Tai},\ and\
  \citenamefont {Wang}}]{Ebadi:2023xhq}%
  \BibitemOpen
  \bibfield  {author} {\bibinfo {author} {\bibfnamefont {R.}~\bibnamefont
  {Ebadi}}, \bibinfo {author} {\bibfnamefont {S.}~\bibnamefont {Kumar}},
  \bibinfo {author} {\bibfnamefont {A.}~\bibnamefont {McCune}}, \bibinfo
  {author} {\bibfnamefont {H.}~\bibnamefont {Tai}}, \ and\ \bibinfo {author}
  {\bibfnamefont {L.-T.}\ \bibnamefont {Wang}},\ }\href {\doibase
  10.1103/PhysRevD.109.083519} {\bibfield  {journal} {\bibinfo  {journal}
  {Phys. Rev. D}\ }\textbf {\bibinfo {volume} {109}},\ \bibinfo {pages}
  {083519} (\bibinfo {year} {2024})},\ \Eprint
  {http://arxiv.org/abs/2307.01248} {arXiv:2307.01248 [astro-ph.CO]}
  \BibitemShut {NoStop}%
\bibitem [{\citenamefont {Figueroa}\ \emph {et~al.}(2024)\citenamefont
  {Figueroa}, \citenamefont {Pieroni}, \citenamefont {Ricciardone},\ and\
  \citenamefont {Simakachorn}}]{Figueroa:2023zhu}%
  \BibitemOpen
  \bibfield  {author} {\bibinfo {author} {\bibfnamefont {D.~G.}\ \bibnamefont
  {Figueroa}}, \bibinfo {author} {\bibfnamefont {M.}~\bibnamefont {Pieroni}},
  \bibinfo {author} {\bibfnamefont {A.}~\bibnamefont {Ricciardone}}, \ and\
  \bibinfo {author} {\bibfnamefont {P.}~\bibnamefont {Simakachorn}},\ }\href
  {\doibase 10.1103/PhysRevLett.132.171002} {\bibfield  {journal} {\bibinfo
  {journal} {Phys. Rev. Lett.}\ }\textbf {\bibinfo {volume} {132}},\ \bibinfo
  {pages} {171002} (\bibinfo {year} {2024})},\ \Eprint
  {http://arxiv.org/abs/2307.02399} {arXiv:2307.02399 [astro-ph.CO]}
  \BibitemShut {NoStop}%
\bibitem [{\citenamefont {Yi}\ \emph {et~al.}(2023)\citenamefont {Yi},
  \citenamefont {Gao}, \citenamefont {Gong}, \citenamefont {Wang},\ and\
  \citenamefont {Zhang}}]{Yi:2023mbm}%
  \BibitemOpen
  \bibfield  {author} {\bibinfo {author} {\bibfnamefont {Z.}~\bibnamefont
  {Yi}}, \bibinfo {author} {\bibfnamefont {Q.}~\bibnamefont {Gao}}, \bibinfo
  {author} {\bibfnamefont {Y.}~\bibnamefont {Gong}}, \bibinfo {author}
  {\bibfnamefont {Y.}~\bibnamefont {Wang}}, \ and\ \bibinfo {author}
  {\bibfnamefont {F.}~\bibnamefont {Zhang}},\ }\href {\doibase
  10.1007/s11433-023-2266-1} {\bibfield  {journal} {\bibinfo  {journal} {Sci.
  China Phys. Mech. Astron.}\ }\textbf {\bibinfo {volume} {66}},\ \bibinfo
  {pages} {120404} (\bibinfo {year} {2023})},\ \Eprint
  {http://arxiv.org/abs/2307.02467} {arXiv:2307.02467 [gr-qc]} \BibitemShut
  {NoStop}%
\bibitem [{\citenamefont {Madge}\ \emph {et~al.}(2023)\citenamefont {Madge},
  \citenamefont {Morgante}, \citenamefont {Puchades-Ib\'a\~nez}, \citenamefont
  {Ramberg}, \citenamefont {Ratzinger}, \citenamefont {Schenk},\ and\
  \citenamefont {Schwaller}}]{Madge:2023dxc}%
  \BibitemOpen
  \bibfield  {author} {\bibinfo {author} {\bibfnamefont {E.}~\bibnamefont
  {Madge}}, \bibinfo {author} {\bibfnamefont {E.}~\bibnamefont {Morgante}},
  \bibinfo {author} {\bibfnamefont {C.}~\bibnamefont {Puchades-Ib\'a\~nez}},
  \bibinfo {author} {\bibfnamefont {N.}~\bibnamefont {Ramberg}}, \bibinfo
  {author} {\bibfnamefont {W.}~\bibnamefont {Ratzinger}}, \bibinfo {author}
  {\bibfnamefont {S.}~\bibnamefont {Schenk}}, \ and\ \bibinfo {author}
  {\bibfnamefont {P.}~\bibnamefont {Schwaller}},\ }\href {\doibase
  10.1007/JHEP10(2023)171} {\bibfield  {journal} {\bibinfo  {journal} {JHEP}\
  }\textbf {\bibinfo {volume} {10}},\ \bibinfo {pages} {171} (\bibinfo {year}
  {2023})},\ \Eprint {http://arxiv.org/abs/2306.14856} {arXiv:2306.14856
  [hep-ph]} \BibitemShut {NoStop}%
\bibitem [{\citenamefont {Firouzjahi}\ and\ \citenamefont
  {Talebian}(2023)}]{Firouzjahi:2023lzg}%
  \BibitemOpen
  \bibfield  {author} {\bibinfo {author} {\bibfnamefont {H.}~\bibnamefont
  {Firouzjahi}}\ and\ \bibinfo {author} {\bibfnamefont {A.}~\bibnamefont
  {Talebian}},\ }\href {\doibase 10.1088/1475-7516/2023/10/032} {\bibfield
  {journal} {\bibinfo  {journal} {JCAP}\ }\textbf {\bibinfo {volume} {10}},\
  \bibinfo {pages} {032} (\bibinfo {year} {2023})},\ \Eprint
  {http://arxiv.org/abs/2307.03164} {arXiv:2307.03164 [gr-qc]} \BibitemShut
  {NoStop}%
\bibitem [{\citenamefont {Wang}\ \emph
  {et~al.}(2024{\natexlab{b}})\citenamefont {Wang}, \citenamefont {Zhao},\ and\
  \citenamefont {Zhu}}]{Wang:2023sij}%
  \BibitemOpen
  \bibfield  {author} {\bibinfo {author} {\bibfnamefont {S.}~\bibnamefont
  {Wang}}, \bibinfo {author} {\bibfnamefont {Z.-C.}\ \bibnamefont {Zhao}}, \
  and\ \bibinfo {author} {\bibfnamefont {Q.-H.}\ \bibnamefont {Zhu}},\ }\href
  {\doibase 10.1103/PhysRevResearch.6.013207} {\bibfield  {journal} {\bibinfo
  {journal} {Phys. Rev. Res.}\ }\textbf {\bibinfo {volume} {6}},\ \bibinfo
  {pages} {013207} (\bibinfo {year} {2024}{\natexlab{b}})},\ \Eprint
  {http://arxiv.org/abs/2307.03095} {arXiv:2307.03095 [astro-ph.CO]}
  \BibitemShut {NoStop}%
\bibitem [{\citenamefont {You}\ \emph {et~al.}(2023)\citenamefont {You},
  \citenamefont {Yi},\ and\ \citenamefont {Wu}}]{You:2023rmn}%
  \BibitemOpen
  \bibfield  {author} {\bibinfo {author} {\bibfnamefont {Z.-Q.}\ \bibnamefont
  {You}}, \bibinfo {author} {\bibfnamefont {Z.}~\bibnamefont {Yi}}, \ and\
  \bibinfo {author} {\bibfnamefont {Y.}~\bibnamefont {Wu}},\ }\href {\doibase
  10.1088/1475-7516/2023/11/065} {\bibfield  {journal} {\bibinfo  {journal}
  {JCAP}\ }\textbf {\bibinfo {volume} {11}},\ \bibinfo {pages} {065} (\bibinfo
  {year} {2023})},\ \Eprint {http://arxiv.org/abs/2307.04419} {arXiv:2307.04419
  [gr-qc]} \BibitemShut {NoStop}%
\bibitem [{\citenamefont {Ye}\ and\ \citenamefont
  {Silvestri}(2024)}]{Ye:2023xyr}%
  \BibitemOpen
  \bibfield  {author} {\bibinfo {author} {\bibfnamefont {G.}~\bibnamefont
  {Ye}}\ and\ \bibinfo {author} {\bibfnamefont {A.}~\bibnamefont {Silvestri}},\
  }\href {\doibase 10.3847/2041-8213/ad2851} {\bibfield  {journal} {\bibinfo
  {journal} {Astrophys. J. Lett.}\ }\textbf {\bibinfo {volume} {963}},\
  \bibinfo {pages} {L15} (\bibinfo {year} {2024})},\ \Eprint
  {http://arxiv.org/abs/2307.05455} {arXiv:2307.05455 [astro-ph.CO]}
  \BibitemShut {NoStop}%
\bibitem [{\citenamefont {Hosseini~Mansoori}\ \emph {et~al.}(2023)\citenamefont
  {Hosseini~Mansoori}, \citenamefont {Felegray}, \citenamefont {Talebian},\
  and\ \citenamefont {Sami}}]{HosseiniMansoori:2023mqh}%
  \BibitemOpen
  \bibfield  {author} {\bibinfo {author} {\bibfnamefont {S.~A.}\ \bibnamefont
  {Hosseini~Mansoori}}, \bibinfo {author} {\bibfnamefont {F.}~\bibnamefont
  {Felegray}}, \bibinfo {author} {\bibfnamefont {A.}~\bibnamefont {Talebian}},
  \ and\ \bibinfo {author} {\bibfnamefont {M.}~\bibnamefont {Sami}},\ }\href
  {\doibase 10.1088/1475-7516/2023/08/067} {\bibfield  {journal} {\bibinfo
  {journal} {JCAP}\ }\textbf {\bibinfo {volume} {08}},\ \bibinfo {pages} {067}
  (\bibinfo {year} {2023})},\ \Eprint {http://arxiv.org/abs/2307.06757}
  {arXiv:2307.06757 [astro-ph.CO]} \BibitemShut {NoStop}%
\bibitem [{\citenamefont {Balaji}\ \emph {et~al.}(2023)\citenamefont {Balaji},
  \citenamefont {Dom\`enech},\ and\ \citenamefont
  {Franciolini}}]{Balaji:2023ehk}%
  \BibitemOpen
  \bibfield  {author} {\bibinfo {author} {\bibfnamefont {S.}~\bibnamefont
  {Balaji}}, \bibinfo {author} {\bibfnamefont {G.}~\bibnamefont {Dom\`enech}},
  \ and\ \bibinfo {author} {\bibfnamefont {G.}~\bibnamefont {Franciolini}},\
  }\href {\doibase 10.1088/1475-7516/2023/10/041} {\bibfield  {journal}
  {\bibinfo  {journal} {JCAP}\ }\textbf {\bibinfo {volume} {10}},\ \bibinfo
  {pages} {041} (\bibinfo {year} {2023})},\ \Eprint
  {http://arxiv.org/abs/2307.08552} {arXiv:2307.08552 [gr-qc]} \BibitemShut
  {NoStop}%
\bibitem [{\citenamefont {Jin}\ \emph {et~al.}(2023)\citenamefont {Jin},
  \citenamefont {Chen}, \citenamefont {Yi}, \citenamefont {You}, \citenamefont
  {Liu},\ and\ \citenamefont {Wu}}]{Jin:2023wri}%
  \BibitemOpen
  \bibfield  {author} {\bibinfo {author} {\bibfnamefont {J.-H.}\ \bibnamefont
  {Jin}}, \bibinfo {author} {\bibfnamefont {Z.-C.}\ \bibnamefont {Chen}},
  \bibinfo {author} {\bibfnamefont {Z.}~\bibnamefont {Yi}}, \bibinfo {author}
  {\bibfnamefont {Z.-Q.}\ \bibnamefont {You}}, \bibinfo {author} {\bibfnamefont
  {L.}~\bibnamefont {Liu}}, \ and\ \bibinfo {author} {\bibfnamefont
  {Y.}~\bibnamefont {Wu}},\ }\href {\doibase 10.1088/1475-7516/2023/09/016}
  {\bibfield  {journal} {\bibinfo  {journal} {JCAP}\ }\textbf {\bibinfo
  {volume} {09}},\ \bibinfo {pages} {016} (\bibinfo {year} {2023})},\ \Eprint
  {http://arxiv.org/abs/2307.08687} {arXiv:2307.08687 [astro-ph.CO]}
  \BibitemShut {NoStop}%
\bibitem [{\citenamefont {Das}\ \emph {et~al.}(2023)\citenamefont {Das},
  \citenamefont {Jaman},\ and\ \citenamefont {Sami}}]{Das:2023nmm}%
  \BibitemOpen
  \bibfield  {author} {\bibinfo {author} {\bibfnamefont {B.}~\bibnamefont
  {Das}}, \bibinfo {author} {\bibfnamefont {N.}~\bibnamefont {Jaman}}, \ and\
  \bibinfo {author} {\bibfnamefont {M.}~\bibnamefont {Sami}},\ }\href {\doibase
  10.1103/PhysRevD.108.103510} {\bibfield  {journal} {\bibinfo  {journal}
  {Phys. Rev. D}\ }\textbf {\bibinfo {volume} {108}},\ \bibinfo {pages}
  {103510} (\bibinfo {year} {2023})},\ \Eprint
  {http://arxiv.org/abs/2307.12913} {arXiv:2307.12913 [gr-qc]} \BibitemShut
  {NoStop}%
\bibitem [{\citenamefont {Afzal}\ \emph {et~al.}(2023)\citenamefont {Afzal}
  \emph {et~al.}}]{NANOGrav:2023hvm}%
  \BibitemOpen
  \bibfield  {author} {\bibinfo {author} {\bibfnamefont {A.}~\bibnamefont
  {Afzal}} \emph {et~al.} (\bibinfo {collaboration} {NANOGrav}),\ }\href
  {\doibase 10.3847/2041-8213/acdc91} {\bibfield  {journal} {\bibinfo
  {journal} {Astrophys. J. Lett.}\ }\textbf {\bibinfo {volume} {951}} (\bibinfo
  {year} {2023}),\ 10.3847/2041-8213/acdc91},\ \Eprint
  {http://arxiv.org/abs/2306.16219} {arXiv:2306.16219 [astro-ph.HE]}
  \BibitemShut {NoStop}%
\bibitem [{\citenamefont {Antoniadis}\ \emph {et~al.}(2024)\citenamefont
  {Antoniadis} \emph {et~al.}}]{EPTA:2023xxk}%
  \BibitemOpen
  \bibfield  {author} {\bibinfo {author} {\bibfnamefont {J.}~\bibnamefont
  {Antoniadis}} \emph {et~al.} (\bibinfo {collaboration} {EPTA, InPTA}),\
  }\href {\doibase 10.1051/0004-6361/202347433} {\bibfield  {journal} {\bibinfo
   {journal} {Astron. Astrophys.}\ }\textbf {\bibinfo {volume} {685}},\
  \bibinfo {pages} {A94} (\bibinfo {year} {2024})},\ \Eprint
  {http://arxiv.org/abs/2306.16227} {arXiv:2306.16227 [astro-ph.CO]}
  \BibitemShut {NoStop}%
\bibitem [{\citenamefont {Drees}\ \emph {et~al.}(2015)\citenamefont {Drees},
  \citenamefont {Hajkarim},\ and\ \citenamefont {Schmitz}}]{Drees:2015exa}%
  \BibitemOpen
  \bibfield  {author} {\bibinfo {author} {\bibfnamefont {M.}~\bibnamefont
  {Drees}}, \bibinfo {author} {\bibfnamefont {F.}~\bibnamefont {Hajkarim}}, \
  and\ \bibinfo {author} {\bibfnamefont {E.~R.}\ \bibnamefont {Schmitz}},\
  }\href {\doibase 10.1088/1475-7516/2015/06/025} {\bibfield  {journal}
  {\bibinfo  {journal} {JCAP}\ }\textbf {\bibinfo {volume} {06}},\ \bibinfo
  {pages} {025} (\bibinfo {year} {2015})},\ \Eprint
  {http://arxiv.org/abs/1503.03513} {arXiv:1503.03513 [hep-ph]} \BibitemShut
  {NoStop}%
\bibitem [{\citenamefont {Saikawa}\ and\ \citenamefont
  {Shirai}(2018)}]{Saikawa:2018rcs}%
  \BibitemOpen
  \bibfield  {author} {\bibinfo {author} {\bibfnamefont {K.}~\bibnamefont
  {Saikawa}}\ and\ \bibinfo {author} {\bibfnamefont {S.}~\bibnamefont
  {Shirai}},\ }\href {\doibase 10.1088/1475-7516/2018/05/035} {\bibfield
  {journal} {\bibinfo  {journal} {JCAP}\ }\textbf {\bibinfo {volume} {1805}},\
  \bibinfo {pages} {035} (\bibinfo {year} {2018})},\ \Eprint
  {http://arxiv.org/abs/1803.01038} {arXiv:1803.01038 [hep-ph]} \BibitemShut
  {NoStop}%
\bibitem [{\citenamefont {Carr}\ \emph {et~al.}(2021)\citenamefont {Carr},
  \citenamefont {Clesse}, \citenamefont {Garc\'\i{}a-Bellido},\ and\
  \citenamefont {K\"uhnel}}]{Carr:2019kxo}%
  \BibitemOpen
  \bibfield  {author} {\bibinfo {author} {\bibfnamefont {B.}~\bibnamefont
  {Carr}}, \bibinfo {author} {\bibfnamefont {S.}~\bibnamefont {Clesse}},
  \bibinfo {author} {\bibfnamefont {J.}~\bibnamefont {Garc\'\i{}a-Bellido}}, \
  and\ \bibinfo {author} {\bibfnamefont {F.}~\bibnamefont {K\"uhnel}},\ }\href
  {\doibase 10.1016/j.dark.2020.100755} {\bibfield  {journal} {\bibinfo
  {journal} {Phys. Dark Univ.}\ }\textbf {\bibinfo {volume} {31}},\ \bibinfo
  {pages} {100755} (\bibinfo {year} {2021})},\ \Eprint
  {http://arxiv.org/abs/1906.08217} {arXiv:1906.08217 [astro-ph.CO]}
  \BibitemShut {NoStop}%
\bibitem [{\citenamefont {Dalianis}\ and\ \citenamefont
  {Kouvaris}(2021)}]{Dalianis:2020gup}%
  \BibitemOpen
  \bibfield  {author} {\bibinfo {author} {\bibfnamefont {I.}~\bibnamefont
  {Dalianis}}\ and\ \bibinfo {author} {\bibfnamefont {C.}~\bibnamefont
  {Kouvaris}},\ }\href {\doibase 10.1088/1475-7516/2021/07/046} {\bibfield
  {journal} {\bibinfo  {journal} {JCAP}\ }\textbf {\bibinfo {volume} {07}},\
  \bibinfo {pages} {046} (\bibinfo {year} {2021})},\ \Eprint
  {http://arxiv.org/abs/2012.09255} {arXiv:2012.09255 [astro-ph.CO]}
  \BibitemShut {NoStop}%
\bibitem [{\citenamefont {Dom\`enech}\ \emph
  {et~al.}(2021{\natexlab{a}})\citenamefont {Dom\`enech}, \citenamefont {Lin},\
  and\ \citenamefont {Sasaki}}]{Domenech:2020ssp}%
  \BibitemOpen
  \bibfield  {author} {\bibinfo {author} {\bibfnamefont {G.}~\bibnamefont
  {Dom\`enech}}, \bibinfo {author} {\bibfnamefont {C.}~\bibnamefont {Lin}}, \
  and\ \bibinfo {author} {\bibfnamefont {M.}~\bibnamefont {Sasaki}},\ }\href
  {\doibase 10.1088/1475-7516/2021/11/E01} {\bibfield  {journal} {\bibinfo
  {journal} {JCAP}\ }\textbf {\bibinfo {volume} {04}},\ \bibinfo {pages} {062}
  (\bibinfo {year} {2021}{\natexlab{a}})},\ \bibinfo {note} {[Erratum: JCAP 11,
  E01 (2021)]},\ \Eprint {http://arxiv.org/abs/2012.08151} {arXiv:2012.08151
  [gr-qc]} \BibitemShut {NoStop}%
\bibitem [{\citenamefont {Lozanov}\ and\ \citenamefont
  {Takhistov}(2023)}]{Lozanov:2022yoy}%
  \BibitemOpen
  \bibfield  {author} {\bibinfo {author} {\bibfnamefont {K.~D.}\ \bibnamefont
  {Lozanov}}\ and\ \bibinfo {author} {\bibfnamefont {V.}~\bibnamefont
  {Takhistov}},\ }\href {\doibase 10.1103/PhysRevLett.130.181002} {\bibfield
  {journal} {\bibinfo  {journal} {Phys. Rev. Lett.}\ }\textbf {\bibinfo
  {volume} {130}},\ \bibinfo {pages} {181002} (\bibinfo {year} {2023})},\
  \Eprint {http://arxiv.org/abs/2204.07152} {arXiv:2204.07152 [astro-ph.CO]}
  \BibitemShut {NoStop}%
\bibitem [{\citenamefont {Bhaumik}\ and\ \citenamefont
  {Jain}(2021)}]{Bhaumik:2020dor}%
  \BibitemOpen
  \bibfield  {author} {\bibinfo {author} {\bibfnamefont {N.}~\bibnamefont
  {Bhaumik}}\ and\ \bibinfo {author} {\bibfnamefont {R.~K.}\ \bibnamefont
  {Jain}},\ }\href {\doibase 10.1103/PhysRevD.104.023531} {\bibfield  {journal}
  {\bibinfo  {journal} {Phys. Rev. D}\ }\textbf {\bibinfo {volume} {104}},\
  \bibinfo {pages} {023531} (\bibinfo {year} {2021})},\ \Eprint
  {http://arxiv.org/abs/2009.10424} {arXiv:2009.10424 [astro-ph.CO]}
  \BibitemShut {NoStop}%
\bibitem [{\citenamefont {Haque}\ \emph {et~al.}(2021)\citenamefont {Haque},
  \citenamefont {Maity}, \citenamefont {Paul},\ and\ \citenamefont
  {Sriramkumar}}]{Haque:2021dha}%
  \BibitemOpen
  \bibfield  {author} {\bibinfo {author} {\bibfnamefont {M.~R.}\ \bibnamefont
  {Haque}}, \bibinfo {author} {\bibfnamefont {D.}~\bibnamefont {Maity}},
  \bibinfo {author} {\bibfnamefont {T.}~\bibnamefont {Paul}}, \ and\ \bibinfo
  {author} {\bibfnamefont {L.}~\bibnamefont {Sriramkumar}},\ }\href {\doibase
  10.1103/PhysRevD.104.063513} {\bibfield  {journal} {\bibinfo  {journal}
  {Phys. Rev. D}\ }\textbf {\bibinfo {volume} {104}},\ \bibinfo {pages}
  {063513} (\bibinfo {year} {2021})},\ \Eprint
  {http://arxiv.org/abs/2105.09242} {arXiv:2105.09242 [astro-ph.CO]}
  \BibitemShut {NoStop}%
\bibitem [{\citenamefont {Papanikolaou}\ \emph {et~al.}(2021)\citenamefont
  {Papanikolaou}, \citenamefont {Vennin},\ and\ \citenamefont
  {Langlois}}]{Papanikolaou:2020qtd}%
  \BibitemOpen
  \bibfield  {author} {\bibinfo {author} {\bibfnamefont {T.}~\bibnamefont
  {Papanikolaou}}, \bibinfo {author} {\bibfnamefont {V.}~\bibnamefont
  {Vennin}}, \ and\ \bibinfo {author} {\bibfnamefont {D.}~\bibnamefont
  {Langlois}},\ }\href {\doibase 10.1088/1475-7516/2021/03/053} {\bibfield
  {journal} {\bibinfo  {journal} {JCAP}\ }\textbf {\bibinfo {volume} {03}},\
  \bibinfo {pages} {053} (\bibinfo {year} {2021})},\ \Eprint
  {http://arxiv.org/abs/2010.11573} {arXiv:2010.11573 [astro-ph.CO]}
  \BibitemShut {NoStop}%
\bibitem [{\citenamefont {Huang}\ \emph {et~al.}(2024)\citenamefont {Huang},
  \citenamefont {Cai}, \citenamefont {Jiang}, \citenamefont {Zhang},\ and\
  \citenamefont {Piao}}]{Huang:2023chx}%
  \BibitemOpen
  \bibfield  {author} {\bibinfo {author} {\bibfnamefont {H.-L.}\ \bibnamefont
  {Huang}}, \bibinfo {author} {\bibfnamefont {Y.}~\bibnamefont {Cai}}, \bibinfo
  {author} {\bibfnamefont {J.-Q.}\ \bibnamefont {Jiang}}, \bibinfo {author}
  {\bibfnamefont {J.}~\bibnamefont {Zhang}}, \ and\ \bibinfo {author}
  {\bibfnamefont {Y.-S.}\ \bibnamefont {Piao}},\ }\href {\doibase
  10.1088/1674-4527/ad683d} {\bibfield  {journal} {\bibinfo  {journal} {Res.
  Astron. Astrophys.}\ }\textbf {\bibinfo {volume} {24}},\ \bibinfo {pages}
  {091001} (\bibinfo {year} {2024})},\ \Eprint
  {http://arxiv.org/abs/2306.17577} {arXiv:2306.17577 [gr-qc]} \BibitemShut
  {NoStop}%
\bibitem [{\citenamefont {Johnson}\ and\ \citenamefont
  {Kamionkowski}(2008)}]{Johnson:2008se}%
  \BibitemOpen
  \bibfield  {author} {\bibinfo {author} {\bibfnamefont {M.~C.}\ \bibnamefont
  {Johnson}}\ and\ \bibinfo {author} {\bibfnamefont {M.}~\bibnamefont
  {Kamionkowski}},\ }\href {\doibase 10.1103/PhysRevD.78.063010} {\bibfield
  {journal} {\bibinfo  {journal} {Phys. Rev. D}\ }\textbf {\bibinfo {volume}
  {78}},\ \bibinfo {pages} {063010} (\bibinfo {year} {2008})},\ \Eprint
  {http://arxiv.org/abs/0805.1748} {arXiv:0805.1748 [astro-ph]} \BibitemShut
  {NoStop}%
\bibitem [{\citenamefont {Turner}(1983)}]{Turner:1983he}%
  \BibitemOpen
  \bibfield  {author} {\bibinfo {author} {\bibfnamefont {M.~S.}\ \bibnamefont
  {Turner}},\ }\href {\doibase 10.1103/PhysRevD.28.1243} {\bibfield  {journal}
  {\bibinfo  {journal} {Phys. Rev. D}\ }\textbf {\bibinfo {volume} {28}},\
  \bibinfo {pages} {1243} (\bibinfo {year} {1983})}\BibitemShut {NoStop}%
\bibitem [{\citenamefont {Poulin}\ \emph {et~al.}(2018)\citenamefont {Poulin},
  \citenamefont {Smith}, \citenamefont {Grin}, \citenamefont {Karwal},\ and\
  \citenamefont {Kamionkowski}}]{Poulin:2018dzj}%
  \BibitemOpen
  \bibfield  {author} {\bibinfo {author} {\bibfnamefont {V.}~\bibnamefont
  {Poulin}}, \bibinfo {author} {\bibfnamefont {T.~L.}\ \bibnamefont {Smith}},
  \bibinfo {author} {\bibfnamefont {D.}~\bibnamefont {Grin}}, \bibinfo {author}
  {\bibfnamefont {T.}~\bibnamefont {Karwal}}, \ and\ \bibinfo {author}
  {\bibfnamefont {M.}~\bibnamefont {Kamionkowski}},\ }\href {\doibase
  10.1103/PhysRevD.98.083525} {\bibfield  {journal} {\bibinfo  {journal} {Phys.
  Rev. D}\ }\textbf {\bibinfo {volume} {98}},\ \bibinfo {pages} {083525}
  (\bibinfo {year} {2018})},\ \Eprint {http://arxiv.org/abs/1806.10608}
  {arXiv:1806.10608 [astro-ph.CO]} \BibitemShut {NoStop}%
\bibitem [{\citenamefont {Vikman}(2005)}]{Vikman:2004dc}%
  \BibitemOpen
  \bibfield  {author} {\bibinfo {author} {\bibfnamefont {A.}~\bibnamefont
  {Vikman}},\ }\href {\doibase 10.1103/PhysRevD.71.023515} {\bibfield
  {journal} {\bibinfo  {journal} {Phys. Rev. D}\ }\textbf {\bibinfo {volume}
  {71}},\ \bibinfo {pages} {023515} (\bibinfo {year} {2005})},\ \Eprint
  {http://arxiv.org/abs/astro-ph/0407107} {arXiv:astro-ph/0407107} \BibitemShut
  {NoStop}%
\bibitem [{\citenamefont {Dom\`enech}\ \emph
  {et~al.}(2021{\natexlab{b}})\citenamefont {Dom\`enech}, \citenamefont
  {Takhistov},\ and\ \citenamefont {Sasaki}}]{Domenech:2021wkk}%
  \BibitemOpen
  \bibfield  {author} {\bibinfo {author} {\bibfnamefont {G.}~\bibnamefont
  {Dom\`enech}}, \bibinfo {author} {\bibfnamefont {V.}~\bibnamefont
  {Takhistov}}, \ and\ \bibinfo {author} {\bibfnamefont {M.}~\bibnamefont
  {Sasaki}},\ }\href {\doibase 10.1016/j.physletb.2021.136722} {\bibfield
  {journal} {\bibinfo  {journal} {Phys. Lett. B}\ }\textbf {\bibinfo {volume}
  {823}},\ \bibinfo {pages} {136722} (\bibinfo {year} {2021}{\natexlab{b}})},\
  \Eprint {http://arxiv.org/abs/2105.06816} {arXiv:2105.06816 [astro-ph.CO]}
  \BibitemShut {NoStop}%
\bibitem [{\citenamefont {Ananda}\ \emph {et~al.}(2007)\citenamefont {Ananda},
  \citenamefont {Clarkson},\ and\ \citenamefont {Wands}}]{Ananda:2006af}%
  \BibitemOpen
  \bibfield  {author} {\bibinfo {author} {\bibfnamefont {K.~N.}\ \bibnamefont
  {Ananda}}, \bibinfo {author} {\bibfnamefont {C.}~\bibnamefont {Clarkson}}, \
  and\ \bibinfo {author} {\bibfnamefont {D.}~\bibnamefont {Wands}},\ }\href
  {\doibase 10.1103/PhysRevD.75.123518} {\bibfield  {journal} {\bibinfo
  {journal} {Phys. Rev.}\ }\textbf {\bibinfo {volume} {D75}},\ \bibinfo {pages}
  {123518} (\bibinfo {year} {2007})},\ \Eprint
  {http://arxiv.org/abs/gr-qc/0612013} {arXiv:gr-qc/0612013 [gr-qc]}
  \BibitemShut {NoStop}%
\bibitem [{\citenamefont {Baumann}\ \emph {et~al.}(2007)\citenamefont
  {Baumann}, \citenamefont {Steinhardt}, \citenamefont {Takahashi},\ and\
  \citenamefont {Ichiki}}]{Baumann:2007zm}%
  \BibitemOpen
  \bibfield  {author} {\bibinfo {author} {\bibfnamefont {D.}~\bibnamefont
  {Baumann}}, \bibinfo {author} {\bibfnamefont {P.~J.}\ \bibnamefont
  {Steinhardt}}, \bibinfo {author} {\bibfnamefont {K.}~\bibnamefont
  {Takahashi}}, \ and\ \bibinfo {author} {\bibfnamefont {K.}~\bibnamefont
  {Ichiki}},\ }\href {\doibase 10.1103/PhysRevD.76.084019} {\bibfield
  {journal} {\bibinfo  {journal} {Phys. Rev.}\ }\textbf {\bibinfo {volume}
  {D76}},\ \bibinfo {pages} {084019} (\bibinfo {year} {2007})},\ \Eprint
  {http://arxiv.org/abs/hep-th/0703290} {arXiv:hep-th/0703290 [hep-th]}
  \BibitemShut {NoStop}%
\bibitem [{\citenamefont {Mollerach}\ \emph {et~al.}(2004)\citenamefont
  {Mollerach}, \citenamefont {Harari},\ and\ \citenamefont
  {Matarrese}}]{Mollerach:2003nq}%
  \BibitemOpen
  \bibfield  {author} {\bibinfo {author} {\bibfnamefont {S.}~\bibnamefont
  {Mollerach}}, \bibinfo {author} {\bibfnamefont {D.}~\bibnamefont {Harari}}, \
  and\ \bibinfo {author} {\bibfnamefont {S.}~\bibnamefont {Matarrese}},\ }\href
  {\doibase 10.1103/PhysRevD.69.063002} {\bibfield  {journal} {\bibinfo
  {journal} {Phys. Rev.}\ }\textbf {\bibinfo {volume} {D69}},\ \bibinfo {pages}
  {063002} (\bibinfo {year} {2004})},\ \Eprint
  {http://arxiv.org/abs/astro-ph/0310711} {arXiv:astro-ph/0310711 [astro-ph]}
  \BibitemShut {NoStop}%
\bibitem [{\citenamefont {Assadullahi}\ and\ \citenamefont
  {Wands}(2010)}]{Assadullahi:2009jc}%
  \BibitemOpen
  \bibfield  {author} {\bibinfo {author} {\bibfnamefont {H.}~\bibnamefont
  {Assadullahi}}\ and\ \bibinfo {author} {\bibfnamefont {D.}~\bibnamefont
  {Wands}},\ }\href {\doibase 10.1103/PhysRevD.81.023527} {\bibfield  {journal}
  {\bibinfo  {journal} {Phys. Rev.}\ }\textbf {\bibinfo {volume} {D81}},\
  \bibinfo {pages} {023527} (\bibinfo {year} {2010})},\ \Eprint
  {http://arxiv.org/abs/0907.4073} {arXiv:0907.4073 [astro-ph.CO]} \BibitemShut
  {NoStop}%
\bibitem [{\citenamefont {Espinosa}\ \emph {et~al.}(2018)\citenamefont
  {Espinosa}, \citenamefont {Racco},\ and\ \citenamefont
  {Riotto}}]{Espinosa:2018eve}%
  \BibitemOpen
  \bibfield  {author} {\bibinfo {author} {\bibfnamefont {J.~R.}\ \bibnamefont
  {Espinosa}}, \bibinfo {author} {\bibfnamefont {D.}~\bibnamefont {Racco}}, \
  and\ \bibinfo {author} {\bibfnamefont {A.}~\bibnamefont {Riotto}},\ }\href
  {\doibase 10.1088/1475-7516/2018/09/012} {\bibfield  {journal} {\bibinfo
  {journal} {JCAP}\ }\textbf {\bibinfo {volume} {09}},\ \bibinfo {pages} {012}
  (\bibinfo {year} {2018})},\ \Eprint {http://arxiv.org/abs/1804.07732}
  {arXiv:1804.07732 [hep-ph]} \BibitemShut {NoStop}%
\bibitem [{\citenamefont {Kohri}\ and\ \citenamefont
  {Terada}(2018)}]{Kohri:2018awv}%
  \BibitemOpen
  \bibfield  {author} {\bibinfo {author} {\bibfnamefont {K.}~\bibnamefont
  {Kohri}}\ and\ \bibinfo {author} {\bibfnamefont {T.}~\bibnamefont {Terada}},\
  }\href {\doibase 10.1103/PhysRevD.97.123532} {\bibfield  {journal} {\bibinfo
  {journal} {Phys. Rev.}\ }\textbf {\bibinfo {volume} {D97}},\ \bibinfo {pages}
  {123532} (\bibinfo {year} {2018})},\ \Eprint
  {http://arxiv.org/abs/1804.08577} {arXiv:1804.08577 [gr-qc]} \BibitemShut
  {NoStop}%
\bibitem [{\citenamefont {Inomata}\ \emph
  {et~al.}(2020{\natexlab{a}})\citenamefont {Inomata}, \citenamefont {Kohri},
  \citenamefont {Nakama},\ and\ \citenamefont {Terada}}]{Inomata:2020tkl}%
  \BibitemOpen
  \bibfield  {author} {\bibinfo {author} {\bibfnamefont {K.}~\bibnamefont
  {Inomata}}, \bibinfo {author} {\bibfnamefont {K.}~\bibnamefont {Kohri}},
  \bibinfo {author} {\bibfnamefont {T.}~\bibnamefont {Nakama}}, \ and\ \bibinfo
  {author} {\bibfnamefont {T.}~\bibnamefont {Terada}},\ }\href {\doibase
  10.1088/1742-6596/1468/1/012002} {\bibfield  {journal} {\bibinfo  {journal}
  {J. Phys. Conf. Ser.}\ }\textbf {\bibinfo {volume} {1468}},\ \bibinfo {pages}
  {012002} (\bibinfo {year} {2020}{\natexlab{a}})}\BibitemShut {NoStop}%
\bibitem [{\citenamefont {Inomata}\ \emph
  {et~al.}(2019{\natexlab{a}})\citenamefont {Inomata}, \citenamefont {Kohri},
  \citenamefont {Nakama},\ and\ \citenamefont {Terada}}]{Inomata:2019ivs}%
  \BibitemOpen
  \bibfield  {author} {\bibinfo {author} {\bibfnamefont {K.}~\bibnamefont
  {Inomata}}, \bibinfo {author} {\bibfnamefont {K.}~\bibnamefont {Kohri}},
  \bibinfo {author} {\bibfnamefont {T.}~\bibnamefont {Nakama}}, \ and\ \bibinfo
  {author} {\bibfnamefont {T.}~\bibnamefont {Terada}},\ }\href {\doibase
  10.1103/PhysRevD.100.043532} {\bibfield  {journal} {\bibinfo  {journal}
  {Phys. Rev. D}\ }\textbf {\bibinfo {volume} {100}},\ \bibinfo {pages}
  {043532} (\bibinfo {year} {2019}{\natexlab{a}})},\ \Eprint
  {http://arxiv.org/abs/1904.12879} {arXiv:1904.12879 [astro-ph.CO]}
  \BibitemShut {NoStop}%
\bibitem [{\citenamefont {Hajkarim}\ and\ \citenamefont
  {Schaffner-Bielich}(2020)}]{Hajkarim:2019nbx}%
  \BibitemOpen
  \bibfield  {author} {\bibinfo {author} {\bibfnamefont {F.}~\bibnamefont
  {Hajkarim}}\ and\ \bibinfo {author} {\bibfnamefont {J.}~\bibnamefont
  {Schaffner-Bielich}},\ }\href {\doibase 10.1103/PhysRevD.101.043522}
  {\bibfield  {journal} {\bibinfo  {journal} {Phys. Rev. D}\ }\textbf {\bibinfo
  {volume} {101}},\ \bibinfo {pages} {043522} (\bibinfo {year} {2020})},\
  \Eprint {http://arxiv.org/abs/1910.12357} {arXiv:1910.12357 [hep-ph]}
  \BibitemShut {NoStop}%
\bibitem [{\citenamefont {Dom\`enech}(2020)}]{Domenech:2019quo}%
  \BibitemOpen
  \bibfield  {author} {\bibinfo {author} {\bibfnamefont {G.}~\bibnamefont
  {Dom\`enech}},\ }\href {\doibase 10.1142/S0218271820500285} {\bibfield
  {journal} {\bibinfo  {journal} {Int. J. Mod. Phys. D}\ }\textbf {\bibinfo
  {volume} {29}},\ \bibinfo {pages} {2050028} (\bibinfo {year} {2020})},\
  \Eprint {http://arxiv.org/abs/1912.05583} {arXiv:1912.05583 [gr-qc]}
  \BibitemShut {NoStop}%
\bibitem [{\citenamefont {Dom\`enech}\ \emph {et~al.}(2020)\citenamefont
  {Dom\`enech}, \citenamefont {Pi},\ and\ \citenamefont
  {Sasaki}}]{Domenech:2020kqm}%
  \BibitemOpen
  \bibfield  {author} {\bibinfo {author} {\bibfnamefont {G.}~\bibnamefont
  {Dom\`enech}}, \bibinfo {author} {\bibfnamefont {S.}~\bibnamefont {Pi}}, \
  and\ \bibinfo {author} {\bibfnamefont {M.}~\bibnamefont {Sasaki}},\ }\href
  {\doibase 10.1088/1475-7516/2020/08/017} {\bibfield  {journal} {\bibinfo
  {journal} {JCAP}\ }\textbf {\bibinfo {volume} {08}},\ \bibinfo {pages} {017}
  (\bibinfo {year} {2020})},\ \Eprint {http://arxiv.org/abs/2005.12314}
  {arXiv:2005.12314 [gr-qc]} \BibitemShut {NoStop}%
\bibitem [{\citenamefont {Dom\`enech}\ and\ \citenamefont
  {Pi}(2022)}]{Domenech:2020ers}%
  \BibitemOpen
  \bibfield  {author} {\bibinfo {author} {\bibfnamefont {G.}~\bibnamefont
  {Dom\`enech}}\ and\ \bibinfo {author} {\bibfnamefont {S.}~\bibnamefont
  {Pi}},\ }\href {\doibase 10.1007/s11433-021-1839-6} {\bibfield  {journal}
  {\bibinfo  {journal} {Sci. China Phys. Mech. Astron.}\ }\textbf {\bibinfo
  {volume} {65}},\ \bibinfo {pages} {230411} (\bibinfo {year} {2022})},\
  \Eprint {http://arxiv.org/abs/2010.03976} {arXiv:2010.03976 [astro-ph.CO]}
  \BibitemShut {NoStop}%
\bibitem [{\citenamefont {Dom\`enech}(2021)}]{Domenech:2021ztg}%
  \BibitemOpen
  \bibfield  {author} {\bibinfo {author} {\bibfnamefont {G.}~\bibnamefont
  {Dom\`enech}},\ }\href {\doibase 10.3390/universe7110398} {\bibfield
  {journal} {\bibinfo  {journal} {Universe}\ }\textbf {\bibinfo {volume} {7}},\
  \bibinfo {pages} {398} (\bibinfo {year} {2021})},\ \Eprint
  {http://arxiv.org/abs/2109.01398} {arXiv:2109.01398 [gr-qc]} \BibitemShut
  {NoStop}%
\bibitem [{\citenamefont {Flauger}\ and\ \citenamefont
  {Weinberg}(2019)}]{Flauger:2019cam}%
  \BibitemOpen
  \bibfield  {author} {\bibinfo {author} {\bibfnamefont {R.}~\bibnamefont
  {Flauger}}\ and\ \bibinfo {author} {\bibfnamefont {S.}~\bibnamefont
  {Weinberg}},\ }\href {\doibase 10.1103/PhysRevD.99.123030} {\bibfield
  {journal} {\bibinfo  {journal} {Phys. Rev. D}\ }\textbf {\bibinfo {volume}
  {99}},\ \bibinfo {pages} {123030} (\bibinfo {year} {2019})},\ \Eprint
  {http://arxiv.org/abs/1906.04853} {arXiv:1906.04853 [hep-th]} \BibitemShut
  {NoStop}%
\bibitem [{\citenamefont {Carr}\ and\ \citenamefont
  {Hawking}(1974)}]{Carr:1974nx}%
  \BibitemOpen
  \bibfield  {author} {\bibinfo {author} {\bibfnamefont {B.~J.}\ \bibnamefont
  {Carr}}\ and\ \bibinfo {author} {\bibfnamefont {S.~W.}\ \bibnamefont
  {Hawking}},\ }\href@noop {} {\bibfield  {journal} {\bibinfo  {journal} {Mon.
  Not. Roy. Astron. Soc.}\ }\textbf {\bibinfo {volume} {168}},\ \bibinfo
  {pages} {399} (\bibinfo {year} {1974})}\BibitemShut {NoStop}%
\bibitem [{\citenamefont {Harada}\ \emph {et~al.}(2016)\citenamefont {Harada},
  \citenamefont {Yoo}, \citenamefont {Kohri}, \citenamefont {Nakao},\ and\
  \citenamefont {Jhingan}}]{Harada:2016mhb}%
  \BibitemOpen
  \bibfield  {author} {\bibinfo {author} {\bibfnamefont {T.}~\bibnamefont
  {Harada}}, \bibinfo {author} {\bibfnamefont {C.-M.}\ \bibnamefont {Yoo}},
  \bibinfo {author} {\bibfnamefont {K.}~\bibnamefont {Kohri}}, \bibinfo
  {author} {\bibfnamefont {K.-i.}\ \bibnamefont {Nakao}}, \ and\ \bibinfo
  {author} {\bibfnamefont {S.}~\bibnamefont {Jhingan}},\ }\href {\doibase
  10.3847/1538-4357/833/1/61} {\bibfield  {journal} {\bibinfo  {journal}
  {Astrophys. J.}\ }\textbf {\bibinfo {volume} {833}},\ \bibinfo {pages} {61}
  (\bibinfo {year} {2016})},\ \Eprint {http://arxiv.org/abs/1609.01588}
  {arXiv:1609.01588 [astro-ph.CO]} \BibitemShut {NoStop}%
\bibitem [{\citenamefont {Nakama}(2020)}]{Nakama:2020kdc}%
  \BibitemOpen
  \bibfield  {author} {\bibinfo {author} {\bibfnamefont {T.}~\bibnamefont
  {Nakama}},\ }\href {\doibase 10.1103/PhysRevD.101.063519} {\bibfield
  {journal} {\bibinfo  {journal} {Phys. Rev. D}\ }\textbf {\bibinfo {volume}
  {101}},\ \bibinfo {pages} {063519} (\bibinfo {year} {2020})}\BibitemShut
  {NoStop}%
\bibitem [{\citenamefont {Inomata}\ \emph
  {et~al.}(2020{\natexlab{b}})\citenamefont {Inomata}, \citenamefont
  {Kawasaki}, \citenamefont {Mukaida}, \citenamefont {Terada},\ and\
  \citenamefont {Yanagida}}]{Inomata:2020lmk}%
  \BibitemOpen
  \bibfield  {author} {\bibinfo {author} {\bibfnamefont {K.}~\bibnamefont
  {Inomata}}, \bibinfo {author} {\bibfnamefont {M.}~\bibnamefont {Kawasaki}},
  \bibinfo {author} {\bibfnamefont {K.}~\bibnamefont {Mukaida}}, \bibinfo
  {author} {\bibfnamefont {T.}~\bibnamefont {Terada}}, \ and\ \bibinfo {author}
  {\bibfnamefont {T.~T.}\ \bibnamefont {Yanagida}},\ }\href {\doibase
  10.1103/PhysRevD.101.123533} {\bibfield  {journal} {\bibinfo  {journal}
  {Phys. Rev. D}\ }\textbf {\bibinfo {volume} {101}},\ \bibinfo {pages}
  {123533} (\bibinfo {year} {2020}{\natexlab{b}})},\ \Eprint
  {http://arxiv.org/abs/2003.10455} {arXiv:2003.10455 [astro-ph.CO]}
  \BibitemShut {NoStop}%
\bibitem [{\citenamefont {Cooke}\ \emph {et~al.}(2014)\citenamefont {Cooke},
  \citenamefont {Pettini}, \citenamefont {Jorgenson}, \citenamefont {Murphy},\
  and\ \citenamefont {Steidel}}]{Cooke:2013cba}%
  \BibitemOpen
  \bibfield  {author} {\bibinfo {author} {\bibfnamefont {R.}~\bibnamefont
  {Cooke}}, \bibinfo {author} {\bibfnamefont {M.}~\bibnamefont {Pettini}},
  \bibinfo {author} {\bibfnamefont {R.~A.}\ \bibnamefont {Jorgenson}}, \bibinfo
  {author} {\bibfnamefont {M.~T.}\ \bibnamefont {Murphy}}, \ and\ \bibinfo
  {author} {\bibfnamefont {C.~C.}\ \bibnamefont {Steidel}},\ }\href {\doibase
  10.1088/0004-637X/781/1/31} {\bibfield  {journal} {\bibinfo  {journal}
  {Astrophys. J.}\ }\textbf {\bibinfo {volume} {781}},\ \bibinfo {pages} {31}
  (\bibinfo {year} {2014})},\ \Eprint {http://arxiv.org/abs/1308.3240}
  {arXiv:1308.3240 [astro-ph.CO]} \BibitemShut {NoStop}%
\bibitem [{\citenamefont {Clarke}\ \emph {et~al.}(2020)\citenamefont {Clarke},
  \citenamefont {Copeland},\ and\ \citenamefont {Moss}}]{Clarke:2020bil}%
  \BibitemOpen
  \bibfield  {author} {\bibinfo {author} {\bibfnamefont {T.~J.}\ \bibnamefont
  {Clarke}}, \bibinfo {author} {\bibfnamefont {E.~J.}\ \bibnamefont
  {Copeland}}, \ and\ \bibinfo {author} {\bibfnamefont {A.}~\bibnamefont
  {Moss}},\ }\href {\doibase 10.1088/1475-7516/2020/10/002} {\bibfield
  {journal} {\bibinfo  {journal} {JCAP}\ }\textbf {\bibinfo {volume} {10}},\
  \bibinfo {pages} {002} (\bibinfo {year} {2020})},\ \Eprint
  {http://arxiv.org/abs/2004.11396} {arXiv:2004.11396 [astro-ph.CO]}
  \BibitemShut {NoStop}%
\bibitem [{\citenamefont {Dalianis}\ and\ \citenamefont
  {Kodaxis}(2022)}]{Dalianis:2021dbs}%
  \BibitemOpen
  \bibfield  {author} {\bibinfo {author} {\bibfnamefont {I.}~\bibnamefont
  {Dalianis}}\ and\ \bibinfo {author} {\bibfnamefont {G.~P.}\ \bibnamefont
  {Kodaxis}},\ }\href {\doibase 10.3390/galaxies10010031} {\bibfield  {journal}
  {\bibinfo  {journal} {Galaxies}\ }\textbf {\bibinfo {volume} {10}},\ \bibinfo
  {pages} {31} (\bibinfo {year} {2022})},\ \Eprint
  {http://arxiv.org/abs/2112.15576} {arXiv:2112.15576 [astro-ph.CO]}
  \BibitemShut {NoStop}%
\bibitem [{\citenamefont {White}\ \emph {et~al.}(2021)\citenamefont {White},
  \citenamefont {Pearce}, \citenamefont {Vagie},\ and\ \citenamefont
  {Kusenko}}]{White:2021hwi}%
  \BibitemOpen
  \bibfield  {author} {\bibinfo {author} {\bibfnamefont {G.}~\bibnamefont
  {White}}, \bibinfo {author} {\bibfnamefont {L.}~\bibnamefont {Pearce}},
  \bibinfo {author} {\bibfnamefont {D.}~\bibnamefont {Vagie}}, \ and\ \bibinfo
  {author} {\bibfnamefont {A.}~\bibnamefont {Kusenko}},\ }\href {\doibase
  10.1103/PhysRevLett.127.181601} {\bibfield  {journal} {\bibinfo  {journal}
  {Phys. Rev. Lett.}\ }\textbf {\bibinfo {volume} {127}},\ \bibinfo {pages}
  {181601} (\bibinfo {year} {2021})},\ \Eprint
  {http://arxiv.org/abs/2105.11655} {arXiv:2105.11655 [hep-ph]} \BibitemShut
  {NoStop}%
\bibitem [{\citenamefont {Kasuya}\ \emph {et~al.}(2023)\citenamefont {Kasuya},
  \citenamefont {Kawasaki},\ and\ \citenamefont {Murai}}]{Kasuya:2022cko}%
  \BibitemOpen
  \bibfield  {author} {\bibinfo {author} {\bibfnamefont {S.}~\bibnamefont
  {Kasuya}}, \bibinfo {author} {\bibfnamefont {M.}~\bibnamefont {Kawasaki}}, \
  and\ \bibinfo {author} {\bibfnamefont {K.}~\bibnamefont {Murai}},\ }\href
  {\doibase 10.1088/1475-7516/2023/05/053} {\bibfield  {journal} {\bibinfo
  {journal} {JCAP}\ }\textbf {\bibinfo {volume} {05}},\ \bibinfo {pages} {053}
  (\bibinfo {year} {2023})},\ \Eprint {http://arxiv.org/abs/2212.13370}
  {arXiv:2212.13370 [astro-ph.CO]} \BibitemShut {NoStop}%
\bibitem [{\citenamefont {Kawasaki}\ and\ \citenamefont
  {Murai}(2024)}]{Kawasaki:2023rfx}%
  \BibitemOpen
  \bibfield  {author} {\bibinfo {author} {\bibfnamefont {M.}~\bibnamefont
  {Kawasaki}}\ and\ \bibinfo {author} {\bibfnamefont {K.}~\bibnamefont
  {Murai}},\ }\href {\doibase 10.1088/1475-7516/2024/01/050} {\bibfield
  {journal} {\bibinfo  {journal} {JCAP}\ }\textbf {\bibinfo {volume} {01}},\
  \bibinfo {pages} {050} (\bibinfo {year} {2024})},\ \Eprint
  {http://arxiv.org/abs/2308.13134} {arXiv:2308.13134 [astro-ph.CO]}
  \BibitemShut {NoStop}%
\bibitem [{\citenamefont {Flores}\ \emph {et~al.}(2023)\citenamefont {Flores},
  \citenamefont {Kusenko}, \citenamefont {Pearce}, \citenamefont
  {Perez-Gonzalez},\ and\ \citenamefont {White}}]{Flores:2023dgp}%
  \BibitemOpen
  \bibfield  {author} {\bibinfo {author} {\bibfnamefont {M.~M.}\ \bibnamefont
  {Flores}}, \bibinfo {author} {\bibfnamefont {A.}~\bibnamefont {Kusenko}},
  \bibinfo {author} {\bibfnamefont {L.}~\bibnamefont {Pearce}}, \bibinfo
  {author} {\bibfnamefont {Y.~F.}\ \bibnamefont {Perez-Gonzalez}}, \ and\
  \bibinfo {author} {\bibfnamefont {G.}~\bibnamefont {White}},\ }\href
  {\doibase 10.1103/PhysRevD.108.123002} {\bibfield  {journal} {\bibinfo
  {journal} {Phys. Rev. D}\ }\textbf {\bibinfo {volume} {108}},\ \bibinfo
  {pages} {123002} (\bibinfo {year} {2023})},\ \Eprint
  {http://arxiv.org/abs/2308.15522} {arXiv:2308.15522 [hep-ph]} \BibitemShut
  {NoStop}%
\bibitem [{\citenamefont {Harigaya}\ \emph {et~al.}(2023)\citenamefont
  {Harigaya}, \citenamefont {Inomata},\ and\ \citenamefont
  {Terada}}]{Harigaya:2023mhl}%
  \BibitemOpen
  \bibfield  {author} {\bibinfo {author} {\bibfnamefont {K.}~\bibnamefont
  {Harigaya}}, \bibinfo {author} {\bibfnamefont {K.}~\bibnamefont {Inomata}}, \
  and\ \bibinfo {author} {\bibfnamefont {T.}~\bibnamefont {Terada}},\ }\href
  {\doibase 10.1103/PhysRevD.108.L081303} {\bibfield  {journal} {\bibinfo
  {journal} {Phys. Rev. D}\ }\textbf {\bibinfo {volume} {108}},\ \bibinfo
  {pages} {L081303} (\bibinfo {year} {2023})},\ \Eprint
  {http://arxiv.org/abs/2305.14242} {arXiv:2305.14242 [hep-ph]} \BibitemShut
  {NoStop}%
\bibitem [{\citenamefont {Pitrou}\ \emph {et~al.}(2013)\citenamefont {Pitrou},
  \citenamefont {Roy},\ and\ \citenamefont {Umeh}}]{Pitrou:2013hga}%
  \BibitemOpen
  \bibfield  {author} {\bibinfo {author} {\bibfnamefont {C.}~\bibnamefont
  {Pitrou}}, \bibinfo {author} {\bibfnamefont {X.}~\bibnamefont {Roy}}, \ and\
  \bibinfo {author} {\bibfnamefont {O.}~\bibnamefont {Umeh}},\ }\href {\doibase
  10.1088/0264-9381/30/16/165002} {\bibfield  {journal} {\bibinfo  {journal}
  {Class. Quant. Grav.}\ }\textbf {\bibinfo {volume} {30}},\ \bibinfo {pages}
  {165002} (\bibinfo {year} {2013})},\ \Eprint {http://arxiv.org/abs/1302.6174}
  {arXiv:1302.6174 [astro-ph.CO]} \BibitemShut {NoStop}%
\bibitem [{\citenamefont {Hawking}(1971)}]{Hawking:1971ei}%
  \BibitemOpen
  \bibfield  {author} {\bibinfo {author} {\bibfnamefont {S.}~\bibnamefont
  {Hawking}},\ }\href@noop {} {\bibfield  {journal} {\bibinfo  {journal} {Mon.
  Not. Roy. Astron. Soc.}\ }\textbf {\bibinfo {volume} {152}},\ \bibinfo
  {pages} {75} (\bibinfo {year} {1971})}\BibitemShut {NoStop}%
\bibitem [{\citenamefont {Khlopov}\ and\ \citenamefont
  {Polnarev}(1980)}]{KHLOPOV1980383}%
  \BibitemOpen
  \bibfield  {author} {\bibinfo {author} {\bibfnamefont {M.}~\bibnamefont
  {Khlopov}}\ and\ \bibinfo {author} {\bibfnamefont {A.}~\bibnamefont
  {Polnarev}},\ }\href {\doibase https://doi.org/10.1016/0370-2693(80)90624-3}
  {\bibfield  {journal} {\bibinfo  {journal} {Physics Letters B}\ }\textbf
  {\bibinfo {volume} {97}},\ \bibinfo {pages} {383} (\bibinfo {year}
  {1980})}\BibitemShut {NoStop}%
\bibitem [{\citenamefont {Khlopov}\ \emph {et~al.}(1985)\citenamefont
  {Khlopov}, \citenamefont {Malomed},\ and\ \citenamefont
  {Zeldovich}}]{Khlopov:1985jw}%
  \BibitemOpen
  \bibfield  {author} {\bibinfo {author} {\bibfnamefont {M.}~\bibnamefont
  {Khlopov}}, \bibinfo {author} {\bibfnamefont {B.~A.}\ \bibnamefont
  {Malomed}}, \ and\ \bibinfo {author} {\bibfnamefont {I.~B.}\ \bibnamefont
  {Zeldovich}},\ }\href@noop {} {\bibfield  {journal} {\bibinfo  {journal}
  {Mon. Not. Roy. Astron. Soc.}\ }\textbf {\bibinfo {volume} {215}},\ \bibinfo
  {pages} {575} (\bibinfo {year} {1985})}\BibitemShut {NoStop}%
\bibitem [{\citenamefont {Carr}\ and\ \citenamefont
  {Kuhnel}(2020)}]{Carr:2020xqk}%
  \BibitemOpen
  \bibfield  {author} {\bibinfo {author} {\bibfnamefont {B.}~\bibnamefont
  {Carr}}\ and\ \bibinfo {author} {\bibfnamefont {F.}~\bibnamefont {Kuhnel}},\
  }\href {\doibase 10.1146/annurev-nucl-050520-125911} {\bibfield  {journal}
  {\bibinfo  {journal} {Ann. Rev. Nucl. Part. Sci.}\ }\textbf {\bibinfo
  {volume} {70}},\ \bibinfo {pages} {355} (\bibinfo {year} {2020})},\ \Eprint
  {http://arxiv.org/abs/2006.02838} {arXiv:2006.02838 [astro-ph.CO]}
  \BibitemShut {NoStop}%
\bibitem [{\citenamefont {Sasaki}\ \emph {et~al.}(2018)\citenamefont {Sasaki},
  \citenamefont {Suyama}, \citenamefont {Tanaka},\ and\ \citenamefont
  {Yokoyama}}]{Sasaki:2018dmp}%
  \BibitemOpen
  \bibfield  {author} {\bibinfo {author} {\bibfnamefont {M.}~\bibnamefont
  {Sasaki}}, \bibinfo {author} {\bibfnamefont {T.}~\bibnamefont {Suyama}},
  \bibinfo {author} {\bibfnamefont {T.}~\bibnamefont {Tanaka}}, \ and\ \bibinfo
  {author} {\bibfnamefont {S.}~\bibnamefont {Yokoyama}},\ }\href {\doibase
  10.1088/1361-6382/aaa7b4} {\bibfield  {journal} {\bibinfo  {journal} {Class.
  Quant. Grav.}\ }\textbf {\bibinfo {volume} {35}},\ \bibinfo {pages} {063001}
  (\bibinfo {year} {2018})},\ \Eprint {http://arxiv.org/abs/1801.05235}
  {arXiv:1801.05235 [astro-ph.CO]} \BibitemShut {NoStop}%
\bibitem [{\citenamefont {Papanikolaou}(2022)}]{Papanikolaou:2022chm}%
  \BibitemOpen
  \bibfield  {author} {\bibinfo {author} {\bibfnamefont {T.}~\bibnamefont
  {Papanikolaou}},\ }\href {\doibase 10.1088/1475-7516/2022/10/089} {\bibfield
  {journal} {\bibinfo  {journal} {JCAP}\ }\textbf {\bibinfo {volume} {10}},\
  \bibinfo {pages} {089} (\bibinfo {year} {2022})},\ \Eprint
  {http://arxiv.org/abs/2207.11041} {arXiv:2207.11041 [astro-ph.CO]}
  \BibitemShut {NoStop}%
\bibitem [{\citenamefont {Basilakos}\ \emph {et~al.}(2024)\citenamefont
  {Basilakos}, \citenamefont {Nanopoulos}, \citenamefont {Papanikolaou},
  \citenamefont {Saridakis},\ and\ \citenamefont
  {Tzerefos}}]{Basilakos:2023xof}%
  \BibitemOpen
  \bibfield  {author} {\bibinfo {author} {\bibfnamefont {S.}~\bibnamefont
  {Basilakos}}, \bibinfo {author} {\bibfnamefont {D.~V.}\ \bibnamefont
  {Nanopoulos}}, \bibinfo {author} {\bibfnamefont {T.}~\bibnamefont
  {Papanikolaou}}, \bibinfo {author} {\bibfnamefont {E.~N.}\ \bibnamefont
  {Saridakis}}, \ and\ \bibinfo {author} {\bibfnamefont {C.}~\bibnamefont
  {Tzerefos}},\ }\href {\doibase 10.1016/j.physletb.2024.138507} {\bibfield
  {journal} {\bibinfo  {journal} {Phys. Lett. B}\ }\textbf {\bibinfo {volume}
  {850}},\ \bibinfo {pages} {138507} (\bibinfo {year} {2024})},\ \Eprint
  {http://arxiv.org/abs/2307.08601} {arXiv:2307.08601 [hep-th]} \BibitemShut
  {NoStop}%
\bibitem [{\citenamefont {Aghanim}\ \emph {et~al.}(2020)\citenamefont {Aghanim}
  \emph {et~al.}}]{Planck:2018vyg}%
  \BibitemOpen
  \bibfield  {author} {\bibinfo {author} {\bibfnamefont {N.}~\bibnamefont
  {Aghanim}} \emph {et~al.} (\bibinfo {collaboration} {Planck}),\ }\href
  {\doibase 10.1051/0004-6361/201833910} {\bibfield  {journal} {\bibinfo
  {journal} {Astron. Astrophys.}\ }\textbf {\bibinfo {volume} {641}},\ \bibinfo
  {pages} {A6} (\bibinfo {year} {2020})},\ \bibinfo {note} {[Erratum:
  Astron.Astrophys. 652, C4 (2021)]},\ \Eprint
  {http://arxiv.org/abs/1807.06209} {arXiv:1807.06209 [astro-ph.CO]}
  \BibitemShut {NoStop}%
\bibitem [{\citenamefont {Agazie}\ \emph
  {et~al.}(2023{\natexlab{b}})\citenamefont {Agazie} \emph
  {et~al.}}]{NANOGrav:2023hfp}%
  \BibitemOpen
  \bibfield  {author} {\bibinfo {author} {\bibfnamefont {G.}~\bibnamefont
  {Agazie}} \emph {et~al.} (\bibinfo {collaboration} {NANOGrav}),\ }\href
  {\doibase 10.3847/2041-8213/ace18b} {\bibfield  {journal} {\bibinfo
  {journal} {Astrophys. J. Lett.}\ }\textbf {\bibinfo {volume} {952}},\
  \bibinfo {pages} {L37} (\bibinfo {year} {2023}{\natexlab{b}})},\ \Eprint
  {http://arxiv.org/abs/2306.16220} {arXiv:2306.16220 [astro-ph.HE]}
  \BibitemShut {NoStop}%
\bibitem [{\citenamefont {Mitridate}(2023)}]{andrea_mitridate_2023}%
  \BibitemOpen
  \bibfield  {author} {\bibinfo {author} {\bibfnamefont {A.}~\bibnamefont
  {Mitridate}},\ }\href {\doibase 10.5281/zenodo.7876430} {\  (\bibinfo {year}
  {2023}),\ 10.5281/zenodo.7876430}\BibitemShut {NoStop}%
\bibitem [{\citenamefont {Mitridate}\ \emph {et~al.}(2023)\citenamefont
  {Mitridate}, \citenamefont {Wright}, \citenamefont {von Eckardstein},
  \citenamefont {Schr\"oder}, \citenamefont {Nay}, \citenamefont {Olum},
  \citenamefont {Schmitz},\ and\ \citenamefont {Trickle}}]{Mitridate:2023oar}%
  \BibitemOpen
  \bibfield  {author} {\bibinfo {author} {\bibfnamefont {A.}~\bibnamefont
  {Mitridate}}, \bibinfo {author} {\bibfnamefont {D.}~\bibnamefont {Wright}},
  \bibinfo {author} {\bibfnamefont {R.}~\bibnamefont {von Eckardstein}},
  \bibinfo {author} {\bibfnamefont {T.}~\bibnamefont {Schr\"oder}}, \bibinfo
  {author} {\bibfnamefont {J.}~\bibnamefont {Nay}}, \bibinfo {author}
  {\bibfnamefont {K.}~\bibnamefont {Olum}}, \bibinfo {author} {\bibfnamefont
  {K.}~\bibnamefont {Schmitz}}, \ and\ \bibinfo {author} {\bibfnamefont
  {T.}~\bibnamefont {Trickle}},\ }\href@noop {} {\  (\bibinfo {year} {2023})},\
  \Eprint {http://arxiv.org/abs/2306.16377} {arXiv:2306.16377 [hep-ph]}
  \BibitemShut {NoStop}%
\bibitem [{\citenamefont {Ellis}\ \emph {et~al.}(2020)\citenamefont {Ellis},
  \citenamefont {Vallisneri}, \citenamefont {Taylor},\ and\ \citenamefont
  {Baker}}]{enterprise}%
  \BibitemOpen
  \bibfield  {author} {\bibinfo {author} {\bibfnamefont {J.~A.}\ \bibnamefont
  {Ellis}}, \bibinfo {author} {\bibfnamefont {M.}~\bibnamefont {Vallisneri}},
  \bibinfo {author} {\bibfnamefont {S.~R.}\ \bibnamefont {Taylor}}, \ and\
  \bibinfo {author} {\bibfnamefont {P.~T.}\ \bibnamefont {Baker}},\ }\href
  {\doibase 10.5281/zenodo.4059815} {\enquote {\bibinfo {title} {Enterprise:
  Enhanced numerical toolbox enabling a robust pulsar inference suite},}\
  }\bibinfo {howpublished} {Zenodo} (\bibinfo {year} {2020})\BibitemShut
  {NoStop}%
\bibitem [{\citenamefont {Lamb}\ \emph {et~al.}(2023)\citenamefont {Lamb},
  \citenamefont {Taylor},\ and\ \citenamefont {van Haasteren}}]{lamb2023rapid}%
  \BibitemOpen
  \bibfield  {author} {\bibinfo {author} {\bibfnamefont {W.~G.}\ \bibnamefont
  {Lamb}}, \bibinfo {author} {\bibfnamefont {S.~R.}\ \bibnamefont {Taylor}}, \
  and\ \bibinfo {author} {\bibfnamefont {R.}~\bibnamefont {van Haasteren}},\
  }\href@noop {} {\bibfield  {journal} {\bibinfo  {journal} {Physical Review
  D}\ }\textbf {\bibinfo {volume} {108}},\ \bibinfo {pages} {103019} (\bibinfo
  {year} {2023})}\BibitemShut {NoStop}%
\bibitem [{\citenamefont {Maggiore}(2018)}]{Maggiore:2018sht}%
  \BibitemOpen
  \bibfield  {author} {\bibinfo {author} {\bibfnamefont {M.}~\bibnamefont
  {Maggiore}},\ }\href@noop {} {\emph {\bibinfo {title} {{Gravitational Waves.
  Vol. 2: Astrophysics and Cosmology}}}}\ (\bibinfo  {publisher} {Oxford
  University Press},\ \bibinfo {year} {2018})\BibitemShut {NoStop}%
\bibitem [{\citenamefont {Bringmann}\ \emph {et~al.}(2023)\citenamefont
  {Bringmann}, \citenamefont {Depta}, \citenamefont {Konstandin}, \citenamefont
  {Schmidt-Hoberg},\ and\ \citenamefont {Tasillo}}]{Bringmann:2023opz}%
  \BibitemOpen
  \bibfield  {author} {\bibinfo {author} {\bibfnamefont {T.}~\bibnamefont
  {Bringmann}}, \bibinfo {author} {\bibfnamefont {P.~F.}\ \bibnamefont
  {Depta}}, \bibinfo {author} {\bibfnamefont {T.}~\bibnamefont {Konstandin}},
  \bibinfo {author} {\bibfnamefont {K.}~\bibnamefont {Schmidt-Hoberg}}, \ and\
  \bibinfo {author} {\bibfnamefont {C.}~\bibnamefont {Tasillo}},\ }\href
  {\doibase 10.1088/1475-7516/2023/11/053} {\bibfield  {journal} {\bibinfo
  {journal} {JCAP}\ }\textbf {\bibinfo {volume} {11}},\ \bibinfo {pages} {053}
  (\bibinfo {year} {2023})},\ \Eprint {http://arxiv.org/abs/2306.09411}
  {arXiv:2306.09411 [astro-ph.CO]} \BibitemShut {NoStop}%
\bibitem [{\citenamefont {Sobotka}\ \emph {et~al.}(2023)\citenamefont
  {Sobotka}, \citenamefont {Erickcek},\ and\ \citenamefont
  {Smith}}]{Sobotka:2022vrr}%
  \BibitemOpen
  \bibfield  {author} {\bibinfo {author} {\bibfnamefont {A.~C.}\ \bibnamefont
  {Sobotka}}, \bibinfo {author} {\bibfnamefont {A.~L.}\ \bibnamefont
  {Erickcek}}, \ and\ \bibinfo {author} {\bibfnamefont {T.~L.}\ \bibnamefont
  {Smith}},\ }\href {\doibase 10.1103/PhysRevD.107.023525} {\bibfield
  {journal} {\bibinfo  {journal} {Phys. Rev. D}\ }\textbf {\bibinfo {volume}
  {107}},\ \bibinfo {pages} {023525} (\bibinfo {year} {2023})},\ \Eprint
  {http://arxiv.org/abs/2207.14308} {arXiv:2207.14308 [astro-ph.CO]}
  \BibitemShut {NoStop}%
\bibitem [{\citenamefont {de~Salas}\ \emph {et~al.}(2015)\citenamefont
  {de~Salas}, \citenamefont {Lattanzi}, \citenamefont {Mangano}, \citenamefont
  {Miele}, \citenamefont {Pastor},\ and\ \citenamefont
  {Pisanti}}]{deSalas:2015glj}%
  \BibitemOpen
  \bibfield  {author} {\bibinfo {author} {\bibfnamefont {P.~F.}\ \bibnamefont
  {de~Salas}}, \bibinfo {author} {\bibfnamefont {M.}~\bibnamefont {Lattanzi}},
  \bibinfo {author} {\bibfnamefont {G.}~\bibnamefont {Mangano}}, \bibinfo
  {author} {\bibfnamefont {G.}~\bibnamefont {Miele}}, \bibinfo {author}
  {\bibfnamefont {S.}~\bibnamefont {Pastor}}, \ and\ \bibinfo {author}
  {\bibfnamefont {O.}~\bibnamefont {Pisanti}},\ }\href {\doibase
  10.1103/PhysRevD.92.123534} {\bibfield  {journal} {\bibinfo  {journal} {Phys.
  Rev. D}\ }\textbf {\bibinfo {volume} {92}},\ \bibinfo {pages} {123534}
  (\bibinfo {year} {2015})},\ \Eprint {http://arxiv.org/abs/1511.00672}
  {arXiv:1511.00672 [astro-ph.CO]} \BibitemShut {NoStop}%
\bibitem [{\citenamefont {Hasegawa}\ \emph {et~al.}(2019)\citenamefont
  {Hasegawa}, \citenamefont {Hiroshima}, \citenamefont {Kohri}, \citenamefont
  {Hansen}, \citenamefont {Tram},\ and\ \citenamefont
  {Hannestad}}]{Hasegawa:2019jsa}%
  \BibitemOpen
  \bibfield  {author} {\bibinfo {author} {\bibfnamefont {T.}~\bibnamefont
  {Hasegawa}}, \bibinfo {author} {\bibfnamefont {N.}~\bibnamefont {Hiroshima}},
  \bibinfo {author} {\bibfnamefont {K.}~\bibnamefont {Kohri}}, \bibinfo
  {author} {\bibfnamefont {R.~S.~L.}\ \bibnamefont {Hansen}}, \bibinfo {author}
  {\bibfnamefont {T.}~\bibnamefont {Tram}}, \ and\ \bibinfo {author}
  {\bibfnamefont {S.}~\bibnamefont {Hannestad}},\ }\href {\doibase
  10.1088/1475-7516/2019/12/012} {\bibfield  {journal} {\bibinfo  {journal}
  {JCAP}\ }\textbf {\bibinfo {volume} {12}},\ \bibinfo {pages} {012} (\bibinfo
  {year} {2019})},\ \Eprint {http://arxiv.org/abs/1908.10189} {arXiv:1908.10189
  [hep-ph]} \BibitemShut {NoStop}%
\bibitem [{\citenamefont {Ade}\ \emph {et~al.}(2016)\citenamefont {Ade} \emph
  {et~al.}}]{Planck:2015fie}%
  \BibitemOpen
  \bibfield  {author} {\bibinfo {author} {\bibfnamefont {P.~A.~R.}\
  \bibnamefont {Ade}} \emph {et~al.} (\bibinfo {collaboration} {Planck}),\
  }\href {\doibase 10.1051/0004-6361/201525830} {\bibfield  {journal} {\bibinfo
   {journal} {Astron. Astrophys.}\ }\textbf {\bibinfo {volume} {594}},\
  \bibinfo {pages} {A13} (\bibinfo {year} {2016})},\ \Eprint
  {http://arxiv.org/abs/1502.01589} {arXiv:1502.01589 [astro-ph.CO]}
  \BibitemShut {NoStop}%
\bibitem [{\citenamefont {Press}\ and\ \citenamefont
  {Schechter}(1974)}]{Press:1973iz}%
  \BibitemOpen
  \bibfield  {author} {\bibinfo {author} {\bibfnamefont {W.~H.}\ \bibnamefont
  {Press}}\ and\ \bibinfo {author} {\bibfnamefont {P.}~\bibnamefont
  {Schechter}},\ }\href {\doibase 10.1086/152650} {\bibfield  {journal}
  {\bibinfo  {journal} {Astrophys. J.}\ }\textbf {\bibinfo {volume} {187}},\
  \bibinfo {pages} {425} (\bibinfo {year} {1974})}\BibitemShut {NoStop}%
\bibitem [{\citenamefont {Ando}\ \emph {et~al.}(2018)\citenamefont {Ando},
  \citenamefont {Inomata},\ and\ \citenamefont {Kawasaki}}]{Ando:2018qdb}%
  \BibitemOpen
  \bibfield  {author} {\bibinfo {author} {\bibfnamefont {K.}~\bibnamefont
  {Ando}}, \bibinfo {author} {\bibfnamefont {K.}~\bibnamefont {Inomata}}, \
  and\ \bibinfo {author} {\bibfnamefont {M.}~\bibnamefont {Kawasaki}},\ }\href
  {\doibase 10.1103/PhysRevD.97.103528} {\bibfield  {journal} {\bibinfo
  {journal} {Phys. Rev.}\ }\textbf {\bibinfo {volume} {D97}},\ \bibinfo {pages}
  {103528} (\bibinfo {year} {2018})},\ \Eprint
  {http://arxiv.org/abs/1802.06393} {arXiv:1802.06393 [astro-ph.CO]}
  \BibitemShut {NoStop}%
\bibitem [{\citenamefont {Oikonomou}(2023)}]{Oikonomou:2023qfz}%
  \BibitemOpen
  \bibfield  {author} {\bibinfo {author} {\bibfnamefont {V.~K.}\ \bibnamefont
  {Oikonomou}},\ }\href {\doibase 10.1103/PhysRevD.108.043516} {\bibfield
  {journal} {\bibinfo  {journal} {Phys. Rev. D}\ }\textbf {\bibinfo {volume}
  {108}},\ \bibinfo {pages} {043516} (\bibinfo {year} {2023})},\ \Eprint
  {http://arxiv.org/abs/2306.17351} {arXiv:2306.17351 [astro-ph.CO]}
  \BibitemShut {NoStop}%
\bibitem [{\citenamefont {Inomata}\ \emph
  {et~al.}(2019{\natexlab{b}})\citenamefont {Inomata}, \citenamefont {Kohri},
  \citenamefont {Nakama},\ and\ \citenamefont {Terada}}]{Inomata:2019zqy}%
  \BibitemOpen
  \bibfield  {author} {\bibinfo {author} {\bibfnamefont {K.}~\bibnamefont
  {Inomata}}, \bibinfo {author} {\bibfnamefont {K.}~\bibnamefont {Kohri}},
  \bibinfo {author} {\bibfnamefont {T.}~\bibnamefont {Nakama}}, \ and\ \bibinfo
  {author} {\bibfnamefont {T.}~\bibnamefont {Terada}},\ }\href {\doibase
  10.1088/1475-7516/2019/10/071} {\bibfield  {journal} {\bibinfo  {journal}
  {JCAP}\ }\textbf {\bibinfo {volume} {10}},\ \bibinfo {pages} {071} (\bibinfo
  {year} {2019}{\natexlab{b}})},\ \bibinfo {note} {[Erratum: JCAP 08, E01
  (2023)]},\ \Eprint {http://arxiv.org/abs/1904.12878} {arXiv:1904.12878
  [astro-ph.CO]} \BibitemShut {NoStop}%
\end{thebibliography}%
\end{multicols}

\end{document}